# A quaternion-based mathematical model for geometrically exact dynamic analysis of cantilevered pipe conveying fluid

Amir Mehdi Dehrouyeh-Semnani[1]

School of Mechanical Engineering, College of Engineering, University of Tehran, Tehran, Iran

## Abstract

For the first time, the development of the nonlinear geometrically exact governing equations and corresponding boundary conditions of hanging cantilevered flexible pipes conveying fluid in the framework of the quaternion system is undertaken. Additionally, the linear model is derived from the nonlinear one for the stability analysis. The linear and nonlinear mathematical formulations based on the geometrically exact rotation-based model are also extracted from the current model. The obtained integro-partial differential-algebraic equations are converted to a set of ordinary differential algebraic equations via the Galerkin discretization technique, and the resulting equations are numerically solved to determine the self-excited oscillation behavior of the system in the post-flutter region. The critical flow velocities, time traces, bifurcation diagrams, phase planes, and deformed configurations are compared with those reported based on the geometrically exact rotation-based model. The comparative studies divulge that the present model can successfully capture the stability and dynamic characteristics of the system. An interesting feature of the current ordinary differential-algebraic equations is that their coefficients are time-independent, unlike those of geometrically exact rotation-based equations in the Galerkin form. The results show this feature leads to a remarkable decrease in the computational cost, although the number of equations increases and the nonlinearity becomes stronger. The challenging issues with the numerical solution utilized in this study are also discussed.

## 1. Introduction

Fluid-conveying pipes supported at both ends undergo buckling instability at critical flow velocities [1-3]. Their nonlinear dynamics in pre- and post-bulking regions based on different approximation models have been extensively investigated [4-12]. But cantilevered pipes conveying fluid become unstable via flutter instability at adequately high flow velocities [3, 13-16]. The nonlinear mathematical formulation derived from the third-order approximation is the well-known model to simulate their nonlinear responses in pre- and post-flutter regions. This approximation model has attracted a significant number of research interests from both theoretical

---

[1] Email addresses: a.m.dehrouye@ut.ac.ir, a.m.dehrouye@gmail.com (A.M. Dehrouyeh-Semnani)

and experimental studies [17-25]. For cantilevered flexible pipes conveying fluid, the third-order approximation model is incapable of estimating their nonlinear behavior when they undergo extremely large deformations. Hence, different nonlinear geometrically exact models [26-28] have been developed to overcome this deficiency of the third-order approximation model. Among them, the geometrically exact rotation-based model has received notable attention. Chen et al. [28] developed this model's governing equation and corresponding boundary conditions for the first time. They studied the post-flutter self-excited oscillations of cantilevered flexible pipes conveying fluid. In another study [29], they employed the geometrically exact rotation-based model to investigate the role of gravity in the nonlinear dynamics of highly flexible cantilevered pipe conveying fluid. Farokhi et al. [30] employed this model to study the extremely large oscillations of cantilevered pipe conveying fluid with added end-mass. In the first step, they conducted a comprehensive study on the self-excited nonlinear oscillations of the system without added end-mass by considering different mass and gravity parameters and an extensive range of dimensionless flow velocities. In the next step, they extensively investigated the role of added end-mass, together with the other effective parameters in the nonlinear dynamics of system. Chen et al. [31] investigated the magnetic control method for large-deformation vibration of cantilevered hard magnetic soft pipes conveying fluid based on the geometrically exact rotation-based model. Dehrouyeh-Semnani [32] employed this model to study the self-excited large-amplitude oscillation of cantilevered hard magnetic soft pipes conveying fluid with uniform and nonuniform magnetizations exposed to an actuating parallel magnetic field in the post-flutter region. Chen et al. [33] developed a geometrically exact rotation-based model for the cantilevered curved pipe conveying fluid and analyzed their nonlinear dynamics with different initial shapes. For the first time, Chen et al. [34] established the geometrically exact rotation-based model for the nonlinear three-dimensional analysis of cantilevered pipe conveying fluid and studied the nonlinear dynamics of system by comparing the new model with the three-dimensional third-order approximation model.

The Galerkin technique was utilized to discretize the partial differential governing equation(s) into a set of ordinary differential equations in the reviewed articles related to the geometrically exact rotation-based model. As reported in these studies, solving these ordinary differential equations has a very high computational cost compared to the third-order approximation model. This is due to the fact that the coefficients in the discretized equation are not constant and are dependent on the rotation, unlike the third-order approximation model. Additionally, for the three-dimensional model, the singularity of the gimbal lock occurs when the pitch angle becomes $\pi/2$. As a solution, the quaternion description method can be utilized. This method was successfully employed to investigate the nonlinear geometrically exact dynamics of three-dimensional beam-type structures [35-39]. It should be pointed out that the main difference between the geometrically exact rotation-based model with the geometrically exact quaternion-based model for dynamic analysis is that the governing equation arises from the rotational-based model is an integro-partial differential equation, but the governing equations arise from the quaternion-based model are integro-partial differential-algebraic equations. Hence, the first aim of this study is that, for the first time, a new mathematical model for the nonlinear geometrically exact pipe conveying fluid in two-dimensional space based on the quaternion description method is developed. The second one focuses on the solution of Galerkin discretized method-based ordinary differential-algebraic equations to lessen the computational cost.

The paper's outline is as follows: In the next section, the problem and geometry are first described. Then, the quaternion-based nonlinear governing equations in terms of integro-partial differential algebraic equations and the associated boundary conditions are derived via the extended version of Hamilton's principle. The linear model for the stability analysis is also derived from the fully nonlinear model. Moreover, the nonlinear and linear rotation-based models are recovered from the current nonlinear and linear quaternion-based models. Finally, the relationship between the oscillation periods of quaternion elements is discussed. In the third section, the obtained equations in the second section are discretized into a set of ordinary differential equations via the Galerkin technique. Additionally, the index of differential-algebraic equations is reduced from 3 to 1 for numerical purposes. In the fourth section, the linear stability and nonlinear geometrically exact dynamic characteristics of system based on the newly developed model are numerically determined and compared with those numerical results reported based on the geometrically exact rotation-based models available in the existing literature. Moreover, the computational cost and solvability of the new mathematical model are discussed. Lastly, the final remarks are presented in the fifth section.

## 2. Mathematical model

### 2.1. Problem description

The system under investigation is a hanging cantilevered flexible pipe of length $L$, cross-sectional area $A$, second moment of area $I$, Young's modulus $E$, mass per unit length $m$, conveying a fluid of mass per unit length $M$ with mean axial velocity $\bar{U}$; see Figure1. The governing equations of motion and the corresponding boundary conditions are established under the following assumptions: (1) the fluid is incompressible but not inviscid; (2) the flow is of constant velocity and free from pulsation; (3) though the strain in the pipe is taken into account small, large deformations are likely (4) the pipe is slender and the nonlinear Euler-Bernoulli beam theory can describe its behavior; (5) the pipe centerline is inextensible; (6) the internal damping is considered, but the external damping is excluded; (7) the pipe motion is planar.

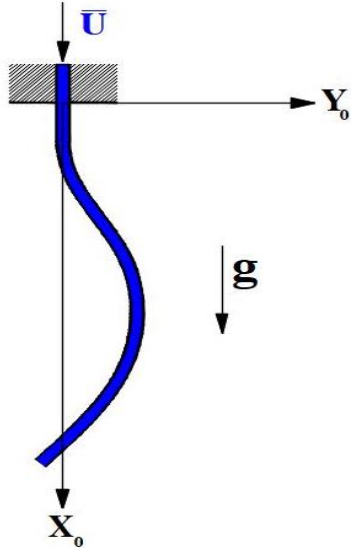

Figure1: Schematic representation of a cantilevered pipe conveying fluid

## 2.2. Geometry description

The Lagrangian coordinate system is considered here as $(X_0, Y_0)$ in which the $X_0$-axis is aligned with the gravity direction. For the cantilevered pipe under consideration ($Y_0 = 0$), the Eulerian coordinate system $(x, y)$ links to the Lagrangian system as follows:

$$x = X_0 + u, \quad y = w, \tag{1}$$

in which $u$ and $w$ stand for the longitudinal and traverse displacements of any point on the pipe, respectively. Defining a curvilinear coordinate, $s^*$, along the pipe length and recalling the fifth assumption, one can obtain $X \equiv s^*$. Therefore, the position vector of any point of the pipe, $\mathbf{r}$, can be represented as follows

$$\mathbf{r}(s^*, t) = x(s^*, t)\mathbf{i} + y(s^*, t)\mathbf{j} = \left(u(s^*, t) + s^*\right)\mathbf{i} + w(s^*, t)\mathbf{j}, \tag{2}$$

in which $t$ is time.

The differential of the variables of $x$ and $y$ in a vector form, $[dx \; dy]^T$, can be written as

$$\begin{bmatrix} dx \\ dy \end{bmatrix} = \mathbf{T} \begin{bmatrix} ds^* \\ 0 \end{bmatrix} = \begin{bmatrix} e_0^2 - e_3^2 & -2e_0 e_3 \\ 2e_0 e_3 & e_0^2 - e_3^2 \end{bmatrix} \begin{bmatrix} ds^* \\ 0 \end{bmatrix} = \begin{bmatrix} (e_0^2 - e_3^2) \\ 2e_0 e_3 \end{bmatrix} ds^*, \tag{3}$$

in which $ds^*$ is the differential of the variable of $s^*$, $\mathbf{T}$ is the rotation matrix, and $e_0(t, s^*)$ and $e_3(t, s^*)$ denote the non-zero quaternion elements in the $X_0 Y_0$ plane. $e_0$ and $e_3$ in terms of the rotation angle, $\theta(t, s^*)$, can be expressed as [35]

$$e_0 = \cos\left(\frac{\theta}{2}\right), \quad e_3 = \sin\left(\frac{\theta}{2}\right), \tag{4}$$

in which

$$\Phi = e_0^2 + e_3^2 = 1. \tag{5}$$

It should be pointed out that the rotation matrix **T** in terms of the rotation angle can be obtained by using Eq. (4) as follows

$$\mathbf{T} = \begin{bmatrix} \cos(\theta) & -\sin(\theta) \\ \sin(\theta) & \cos(\theta) \end{bmatrix}. \tag{6}$$

According to Eq. (3), the elements of Eulerian coordinate system, $(x, y)$, based on the elements of quaternion system, $(e_0, e_3)$, can be formulated as

$$x = \int_0^{s^*} \left(e_0^2 - e_3^2\right) d\hat{s}, \quad y = \int_0^{s^*} 2e_0 e_3 \, d\hat{s}. \tag{7}$$

When the $X_0 Y_0$ coordinate system is rotated by the rotation angle $\theta$ about the $Z_0$ axis to the coordinate system $xy$; the unit vectors of these two coordinate systems are related by [35]

$$\begin{bmatrix} \mathbf{i} \\ \mathbf{j} \end{bmatrix} = \mathbf{T}^* \begin{bmatrix} \mathbf{i}_0 \\ \mathbf{j}_0 \end{bmatrix} = \begin{bmatrix} e_0^2 - e_3^2 & 2e_0 e_3 \\ -2e_0 e_3 & e_0^2 - e_3^2 \end{bmatrix} \begin{bmatrix} \mathbf{i}_0 \\ \mathbf{j}_0 \end{bmatrix}, \tag{8}$$

in which **i** and **j** represent the unit vectors along the $x$ and $y$ axes, and $\mathbf{i}_0$ and $\mathbf{j}_0$ stand for the unit vector along the $X_0$ and $Y_0$ axes. In view of Eq. (4), matrix $\mathbf{T}^*$ in terms of the rotation angle can be expressed as [40]

$$\mathbf{T}^* = \begin{bmatrix} \cos(\theta) & \sin(\theta) \\ -\sin(\theta) & \cos(\theta) \end{bmatrix}. \tag{9}$$

It should be noticed that matrices **T** and $\mathbf{T}^*$ are not the same.

### 2.3. Energy method

The curvature of the pipe, $\kappa$, can be formulated as [40]

$$\kappa = \mathbf{i}'.\mathbf{j} = -\mathbf{j}'.\mathbf{i}, \tag{10}$$

in which the prime symbol denotes the derivative with respect to $s^*$. In view of Eqs. (10) and (8), the curvature as functions of the quaternion elements can be written as

$$\kappa = \left(\left(e_0^2 - e_3^2\right)' \mathbf{i}_0 + 2e_0 e_3 \mathbf{j}_0\right).\left(-2e_0 e_3 \mathbf{i}_0 + \left(e_0^2 - e_3^2\right) \mathbf{j}_0\right) = -2\left(e_0' e_3 - e_0 e_3'\right). \tag{11}$$

It should be noted that the curvature reported by Chen et al. [41] is equal to $\kappa = 2\left(e_0' e_3 - e_0 e_3'\right)$. It is due to the fact that they used **T** rather than $\mathbf{T}^*$ in Eq. (8). Additionally, it is worth noting that one can obtain $\kappa = \theta'$ by substituting Eq. (4) into Eq. (11).

According to the Euler-Bernoulli beam theory, the non-zero components of strain and stress tensors can be obtained by

$$\varepsilon_{xx}(t,s^*) = -z\kappa = 2z\left(e'_0 e_3 - e_0 e'_3\right), \tag{12}$$

$$\sigma_{xx}(t,s^*) = E\varepsilon_{xx} = 2Ez\left(e'_0 e_3 - e_0 e'_3\right). \tag{13}$$

In view of Eqs. (12) and (13), the variation form of the potential energy of the pipe, $\delta U_p$, can be formulated as [41]

$$\begin{aligned}
\delta U_p(t) &= \delta \int_V \sigma_{xx}\varepsilon_{xx}dV = EI\int_0^L \delta\kappa^2 ds^* = EI\int_0^L \delta\left(-2\left(e'_0 e_3 - e_0 e'_3\right)\right)^2 ds^* = \\
&4EI\Bigg(\int_0^L\bigg[\left(-e''_0 e_3^2 + e_0 e_3 e''_3 + 2e_0\left(e'_3\right)^2 - 2e'_0 e_3 e'_3\right)\delta e_0 + \\
&\quad\left(-e_0^2 e''_3 + e_0 e_3 e''_0 + 2\left(e'_0\right)^2 e_3 - 2e_0 e'_0 e'_3\right)\delta e_3\bigg]ds^* + \\
&\quad\left(e'_0 e_3^2 - e_0 e'_3 e_3\right)\delta e_0\bigg|_0^L + \left(e'_3 e_0^2 - e_0 e'_0 e_3\right)\delta e_3\bigg|_0^L\Bigg),
\end{aligned} \tag{14}$$

in which $V$ is the volume of pipe.

It should be mentioned that the variation form of potential energy presented in Eq. (14) is equal to that reported by Chen et al. [41]. It is due to the fact that in the calculation of the potential energy, $\kappa^2$ is utilized.

The variation form of the potential energy due to the gravity, $\delta U_g$, can be formulated as [41]

$$\begin{aligned}
\delta U_g &= -\int_0^L (M+m)g\delta x ds = -\int_0^L (M+m)g\left(\delta\int_0^{s^*}\left(e_0^2 - e_3^2\right)d\hat{s}\right)ds^* = \\
&(M+m)g\int_0^L (L-s^*)\left(2e_3\delta e_3 - 2e_0\delta e_0\right)ds^*.
\end{aligned} \tag{15}$$

In the framework of the Euler-Bernoulli beam theory and the Kelvin-Voight model, the virtual work due to the structural damping of pipe, $\delta W_d$, can be written as

$$\begin{aligned}
\delta W_d(t) &= -E\alpha^* I\int_0^L \left(\dot{\kappa}\delta\kappa\right)ds^* = -\alpha^* I\int_0^L 4\left(\dot{e}'_0 e_3 + e'_0\dot{e}_3 - \dot{e}_0 e'_3 - e_0\dot{e}'_3\right)\delta\left(e'_0 e_3 - e_0 e'_3\right)ds = \\
&-\alpha^* I\int_0^L\Bigg(\bigg[-8\left(e'_3\dot{e}'_0 e_3 + e'_3 e'_0\dot{e}_3 - e'_3\dot{e}_0 e'_3 - e'_3 e_0\dot{e}'_3\right) - 4\left(e_3\dot{e}'_0 e'_3 + e_3\dot{e}''_0 e_3 + e_3 e'_0\dot{e}'_3 + e_3 e''_0\dot{e}_3 - e_3\dot{e}_0 e''_3 - \\
&\quad e_3\dot{e}'_0 e'_3 - e_3 e_0\dot{e}''_3 - e_3 e'_0\dot{e}'_3\right)\bigg]\delta e_0 + \bigg[8\left(e'_0\dot{e}'_0 e_3 + e'_0 e'_0\dot{e}_3 - e'_0\dot{e}_0 e'_3 - e'_0 e_0\dot{e}'_3\right) + \\
&\quad 4\left(e_0\dot{e}'_0 e'_3 + e_0\dot{e}''_0 e_3 + e_0 e'_0\dot{e}'_3 + e_0 e''_0\dot{e}_3 - e_0\dot{e}_0 e''_3 - e_0\dot{e}'_0 e'_3 - e_0 e_0\dot{e}''_3 - e_0 e'_0\dot{e}'_3\right)\bigg]\delta e_3\Bigg)ds,
\end{aligned} \tag{16}$$

in which $\alpha^*$ denotes the damping coefficient and the dot symbol represents the derivative with respect to $t$.

The velocity of an infinitesimal element of the pipe, $\mathbf{V}_p$, and the fluid, $\mathbf{V}_f$, respectively, can be expressed as

$$\mathbf{V}_p = \dot{\mathbf{r}} = \dot{x}\mathbf{i} + \dot{y}\mathbf{j} = \left(\int_0^s \left(2e_0\dot{e}_0 - 2e_3\dot{e}_3\right)ds'\right)\mathbf{i} + \left(\int_0^s 2\left(\dot{e}_0 e_3 + e_0\dot{e}_3\right)ds'\right)\mathbf{j},$$

$$\mathbf{V}_f = \dot{\mathbf{r}} + \bar{U}\mathbf{r}' = \left(\int_0^s \left(2e_0\dot{e}_0 - 2e_3\dot{e}_3\right)ds' + \bar{U}\left(e_0^2 - e_3^2\right)\right)\mathbf{i} + \left(\int_0^s 2\left(\dot{e}_0 e_3 + e_0\dot{e}_3\right)ds' + \bar{U}\left(2e_0 e_3\right)\right)\mathbf{j}.$$

(17)

The kinetic energy of the system, $T$, can be obtained by

$$T(t) = \frac{1}{2}m\int_0^L \mathbf{V}_p^{\;2}ds + \frac{1}{2}M\int_0^L \mathbf{V}_f^{\;2}ds =$$

$$\frac{1}{2}(M+m)\int_0^L \left(\left(\int_0^s \left(2e_0\dot{e}_0 - 2e_3\dot{e}_3\right)ds'\right)^2 + \left(\int_0^s 2\left(\dot{e}_0 e_3 + e_0\dot{e}_3\right)ds'\right)^2\right)ds +$$

$$\frac{1}{2}M\int_0^L \left(\bar{U}^2\left(e_0^2 - e_3^2\right)^2 + \bar{U}^2\left(2e_0 e_3\right)^2\right)ds +$$

$$\frac{1}{2}M\int_0^L \left(2\bar{U}\left(e_0^2 - e_3^2\right)\int_0^s \left(2e_0\dot{e}_0 - 2e_3\dot{e}_3\right)ds'\right)ds.$$

(18)

Considering the following relationship

$$\frac{1}{2}M\int_0^L \left(\bar{U}^2\left(e_0^2 - e_3^2\right)^2 + \bar{U}^2\left(2e_0 e_3\right)^2\right)ds = \frac{1}{2}M\int_0^L \bar{U}^2\left(\underbrace{e_0^2 + e_3^2}_{=1}\right)^2 ds = \frac{1}{2}M\int_0^L \bar{U}^2 ds,$$

(19)

the variation form of the kinetic energy can be written as

$$\delta T(t) = (M+m)\int_0^L \left( \left[ 2e_0 \int_L^{s^*} \int_0^{\hat{s}} 2\left(\ddot{e}_0 e_0 + (\dot{e}_0)^2 - \ddot{e}_3 e_3 - (\dot{e}_3)^2\right) d\tilde{s}d\hat{s} + \right. \right.$$

$$\left. 2e_3 \int_L^{s^*} \int_0^{\hat{s}} 2\left(e_0\ddot{e}_3 + 2\dot{e}_0\dot{e}_3 + \ddot{e}_0 e_3\right) d\tilde{s}d\hat{s} \right] \delta e_0 +$$

$$\left[ 2e_3 \int_L^{s^*} \int_0^{\hat{s}} 2\left(\ddot{e}_0 e_0 + (\dot{e}_0)^2 - \ddot{e}_3 e_3 - (\dot{e}_3)^2\right) d\tilde{s}d\hat{s} + \right.$$

$$\left. \left. 2e_0 \int_L^{s^*} \int_0^{\hat{s}} 2\left(e_0\ddot{e}_3 + 2\dot{e}_0\dot{e}_3 + \ddot{e}_0 e_3\right) d\tilde{s}d\hat{s} \right] \delta e_3 \right) ds^* +$$

$$M\bar{U}\int_0^L \left( \left[ \left( 2e_0 \int_L^{s^*} 2\left(e_0\dot{e}_0 - e_3\dot{e}_3\right) d\hat{s} \right) + \left( 2e_0 \int_L^{s^*} 2\left(e_0\dot{e}_0 - e_3\dot{e}_3\right) d\hat{s} \right) \right. \right.$$

$$\left. + \left( 2e_3 \int_L^{s^*} 2\left(\dot{e}_0 e_3 + e_0\dot{e}_3\right) d\hat{s} \right) + \left( 2e_3 \int_L^{s^*} 2\left(\dot{e}_0 e_3 + e_0\dot{e}_3\right) d\hat{s} \right) \right] \delta e_0 \quad (20)$$

$$+ \left( -2e_3 \int_L^{s^*} 2\left(e_0\dot{e}_0 - e_3\dot{e}_3\right) d\hat{s} \right) + \left( -2e_3 \int_L^{s^*} 2\left(e_0\dot{e}_0 - e_3\dot{e}_3\right) d\hat{s} \right)$$

$$\left. + \left( 2e_0 \int_L^{s^*} 2\left(\dot{e}_0 e_3 + e_0\dot{e}_3\right) d\hat{s} \right) + \left( 2e_0 \int_L^{s^*} 2\left(\dot{e}_0 e_3 + e_0\dot{e}_3\right) d\hat{s} \right) \right] \delta e_3 \right) ds^*$$

$$+ \underbrace{\left( M\bar{U}\int_0^{s^*} 2\left(e_0\dot{e}_0 - e_3\dot{e}_3\right) d\hat{s}\,\delta \int_0^{s^*} \left(e_0^2 - e_3^2\right) d\hat{s} \right)\Bigg|_{s=0}^{s=L}}_{\mathbf{A}}$$

$$+ \underbrace{\left( M\bar{U}\int_0^{s^*} 2\left(\dot{e}_0 e_3 + e_0\dot{e}_3\right) d\hat{s}\,\delta \int_0^{s^*} \left(2e_0 e_3\right) d\hat{s} \right)\Bigg|_{s=0}^{s=L}}_{\mathbf{B}}.$$

It should be pointed out that to derive Eq. (20) from Eq. (18), the following formulation given by Semler et al. [19] is utilized.

$$\int_0^L g(s^*)\left( \int_0^{s^*} f(\hat{s})\delta e_i d\hat{s} \right) ds^* = -\int_0^L \left( \int_L^{s^*} g(\hat{s}) d\hat{s} \right) f(s^*)\delta e_i ds^*. \quad (21)$$

The virtual work done by the fluid exiting the pipe, $\delta W_f$, can be obtained by

$$\int_{t_1}^{t_2} \delta W_f(t)dt = -\int_{t_1}^{t_2}\left[M\bar{U}(\dot{\mathbf{r}}_L + U\boldsymbol{\tau}_L)\delta\mathbf{r}_L\right]dt =$$

$$-\int_{t_1}^{t_2} M\bar{U}\left[(\dot{x}(L,t) + \bar{U}x'(L,t))\delta x(L,t) + (\dot{y}(L,t) + Uy'(L,t))\delta y(L,t)\right]dt =$$

$$-\int_{t_1}^{t_2} M\bar{U}\left[\left(\int_0^L 2(e_0\dot{e}_0 - e_3\dot{e}_3)ds' + \bar{U}(e_0^2(L) - e_3^2(L))\right)\left(\int_0^L \delta(e_0^2 - e_3^2)ds^*\right)\right.$$

$$\left. + \left(\int_0^L 2(\dot{e}_0 e_3 + e_0\dot{e}_3)ds' + \bar{U}(2e_0(L)e_3(L))\right)\int_0^L \delta(2e_0 e_3)ds^*\right]dt =$$

$$\int_{t_1}^{t_2}\left[M\bar{U}^2\int_0^L\left(-\left[2e_0\left(e_0^2(L) - e_3^2(L)\right) + 2e_3\left(2e_0(L)e_3(L)\right)\right]\delta e_0\right.\right. \tag{22}$$

$$\left. -\left[-2e_3\left(e_0^2(L) - e_3^2(L)\right) + 2e_0\left(2e_0(L)e_3(L)\right)\right]\delta e_3\right)ds^*$$

$$\underbrace{-\left(M\bar{U}\int_0^L 2(e_0\dot{e}_0 - e_3\dot{e}_3)ds^*\delta\int_0^L(e_0^2 - e_3^2)ds^*\right)}_{\mathbf{A}}$$

$$\underbrace{-\left(M\bar{U}\int_0^L 2(\dot{e}_0 e_3 + e_0\dot{e}_3)ds^*\delta\int_0^L(2e_0 e_3)ds^*\right)}_{\mathbf{B}}\right]dt.$$

in which $\mathbf{r}_L$ and $\boldsymbol{\tau}_L$ denote the position and tangential unit vectors at the end of the pipe, respectively. It is worth noting that the original formulation of $\delta W_f$ was proposed by Benjamin [42].

### 2.4. Nonlinear formulation

The governing equations and corresponding boundary conditions of the system can be obtained by employing the extended version of Hamilton's principle as follows

$$\int_{t_1}^{t_2}\left(\delta T - \delta U_p - \delta U_g\right)dt + \int_{t_1}^{t_2}\left(\delta W_f + \delta W_d\right)dt - \int_{t_1}^{t_2}\delta\int_V \bar{\lambda}\Phi dV = 0, \tag{23}$$

in which $\bar{\lambda}(s^*, t)$ is the Lagrangian multiplier function. Before deriving the governing equations and associated boundary conditions, it should be pointed out that the terms **A** and **B** in Eq. (22) cancel the corresponding terms in Eq. (20).

Substituting Eqs. (14), (15), (16), (20), and (22) into Eq. (23) and utilizing the following non-dimensional quantities

$$s = \frac{s^*}{L}, \ (\tau, \alpha) = \sqrt{\frac{EI}{(M+m)L^4}}(t, \alpha^*), \ U = \left(\frac{M}{EI}\right)^{1/2}\bar{U},$$

$$\beta = \frac{M}{M+m}, \ \gamma = \frac{M+m}{EI}gL^3, \ \lambda = \frac{\bar{\lambda}AL^2}{EI}, \tag{24}$$

yields

$$4\left(e_0'' e_3^2 - e_0 e_3 e_3'' - 2e_0 (e_3')^2 + 2e_0' e_3 e_3'\right) + \gamma(2e_0)(1-s)$$

$$+\left[2e_0 \int_1^s \int_0^{\hat{s}} 2\left(e_0 \ddot{e}_0 + (\dot{e}_0)^2 - e_3 \ddot{e}_3 - (\dot{e}_3)^2\right) d\tilde{s} d\hat{s}\right]$$

$$+\left[2e_3 \int_1^s \int_0^{\hat{s}} 2\left(e_0 \ddot{e}_3 + 2\dot{e}_0 \dot{e}_3 + e_3 \ddot{e}_0\right) d\tilde{s} d\hat{s}\right]$$

$$+2U\sqrt{\beta}\left[2e_0 \int_1^s 2\left(e_0 \dot{e}_0 - e_3 \dot{e}_3\right) d\hat{s}\right]$$

$$+2U\sqrt{\beta}\left[2e_3 \int_1^s 2\left(e_3 \dot{e}_0 + \dot{e}_3 e_0\right) d\hat{s}\right] \quad (25)$$

$$-U^2\left[2e_0\left(e_0^2(1) - e_3^2(1)\right) + 2e_3\left(2e_0(1)e_3(1)\right)\right]$$

$$-\alpha\left[-8(e_3' e_3 \dot{e}_0' + e_3' e_0' \dot{e}_3 - e_3' e_3' \dot{e}_0 - e_3' e_0 \dot{e}_3') - \right.$$

$$4(e_3 e_3' \dot{e}_0' + e_3 e_3 \dot{e}_0'' + e_3 e_0' \dot{e}_3' + e_3 e_0'' \dot{e}_3 - e_3 e_3'' \dot{e}_0 -$$

$$\left. e_3 e_3' \dot{e}_0' - e_3 e_0 \dot{e}_3'' - e_3 e_0' \dot{e}_3')\right] - 2e_0 \lambda = 0,$$

$$4\left(e_0^2 e_3'' - e_0 e_0'' e_3 - 2(e_0')^2 e_3 + 2e_0 e_0' e_3'\right) - \gamma(2e_3)(1-s)$$

$$-\left[2e_3 \int_1^s \int_0^{\hat{s}} 2\left(e_0 \ddot{e}_0 + (\dot{e}_0)^2 - e_3 \ddot{e}_3 - (\dot{e}_3)^2\right) d\tilde{s} d\hat{s}\right]$$

$$+\left[2e_0 \int_1^s \int_0^{\hat{s}} 2\left(e_0 \ddot{e}_3 + 2\dot{e}_0 \dot{e}_3 + e_3 \ddot{e}_0\right) d\tilde{s} d\hat{s}\right]$$

$$-2U\sqrt{\beta}\left[2e_3 \int_1^s 2\left(e_0 \dot{e}_0 - e_3 \dot{e}_3\right) d\hat{s}\right]$$

$$+2U\sqrt{\beta}\left[2e_0 \int_1^s 2\left(e_3 \dot{e}_0 + \dot{e}_3 e_0\right) d\hat{s}\right] \quad (26)$$

$$-U^2\left[-2e_3\left(e_0^2(1) - e_3^2(1)\right) + 2e_0\left(2e_0(1)e_3(1)\right)\right]$$

$$-\alpha\left[8(e_0' e_3 \dot{e}_0' + e_0' e_0' \dot{e}_3 - e_0' e_3' \dot{e}_0 - e_0' e_0 \dot{e}_3') + \right.$$

$$4(e_0 e_3' \dot{e}_0' + e_0 e_3 \dot{e}_0'' + e_0 e_0' \dot{e}_3' + e_0 e_0'' \dot{e}_3 -$$

$$\left. e_0 e_3'' \dot{e}_0 - e_0 e_3' \dot{e}_0' - e_0 e_0 \dot{e}_3'' - e_0 e_0' \dot{e}_3')\right] - 2\lambda e_3 = 0,$$

$$e_0^2 + e_3^2 = 1, \quad (27)$$

$$e_0' e_3^2 - e_0 e_3' e_3 = 0 \text{ or } e_0 = \bar{e}_0 \text{ at } \zeta=0 \ \& \ 1,$$
$$e_3' e_0^2 - e_0 e_0' e_3 = 0 \text{ or } e_3 = \bar{e}_3 \text{ at } \zeta=0 \ \& \ 1, \quad (28)$$

where Eq. (28) can be rewritten as

$$e_0(0) = 1, \ e_3(0) = 0, \ e_0'(1) = 0, \ e_3'(1) = 0. \tag{29}$$

in which $s$ is the non-dimensional distance along the pipe, $\tau$ is the non-dimensional time, $\alpha$ is the non-dimensional structural damping coefficient, $U$ is the non-dimensional flow velocity, $\beta$ is a mass parameter, $\gamma$ is the non-dimensional gravity parameter, and $\lambda$ is the non-dimensional Lagrangian multiplier function. The non-dimensional form of the longitudinal and transverse displacements, respectively, can be formulated as

$$\zeta = \frac{u}{L} = \int_0^s \left( e_0^2 - e_3^2 \right) d\hat{s} - 1, \ \eta = \frac{w}{L} = \int_0^s 2 e_0 e_3 \ d\hat{s}. \tag{30}$$

## 2.5. Linear formulation

To obtain the linear form of mathematical formulation, one can assume $e_0 = \cos(\theta/2) \approx 1$. Applying this assumption to Eq. (25) and neglecting the nonlinear terms result in

$$2\gamma(1-s) - 2U^2 - 2\lambda = 0. \tag{31}$$

Applying the aforementioned assumption to Eq. (26) gives

$$4e_3'' - \gamma(2e_3)(1-s) + \left[ 2\int_1^s \int_0^{\hat{s}} 2(\ddot{e}_3) d\hat{s} d\hat{s} \right] + 2U\sqrt{\beta} \left[ 2\int_1^s 2(\dot{e}_3) d\hat{s} \right] \\
- U^2 \left[ -2e_3 + 2(2e_3(1)) \right] - 4\alpha \dot{e}_3'' - 2\lambda e_3 = 0. \tag{32}$$

Substituting $\lambda$ from Eq.(31) into Eq. (32) gives the linear form of governing equations as follows

$$4e_3'' - 2\gamma(2e_3)(1-s) + \left[ 2\int_1^s \int_0^{\hat{s}} 2(\ddot{e}_3) d\hat{s} d\hat{s} \right] + 2U\sqrt{\beta} \left[ 2\int_1^s 2(\dot{e}_3) d\hat{s} \right] \\
- 4\alpha \dot{e}_3'' - 2U^2(-2e_3) - U^2 \left[ 2(2e_3(1)) \right] = 0. \tag{33}$$

Moreover, in view of Eq. (29), the boundary conditions of the system in the linear and nonlinear forms are the same.

## 2.6. Extracting the rotation-based model from the quaternion-based model

Both the linear and nonlinear governing equations and boundary conditions of the system based on the rotation angle can be recovered from the current formulations based on the quaternion system.

In view of Eq. (4), the derivatives of quaternion elements in terms of rotation angle and its derivatives can be formulated as

$$e_0' = -\frac{\theta'}{2}\sin\left(\frac{\theta}{2}\right), \ e_3' = \frac{\theta'}{2}\cos\left(\frac{\theta}{2}\right), \ \dot{e}_0 = -\frac{\dot{\theta}}{2}\sin\left(\frac{\theta}{2}\right), \ \dot{e}_3 = \frac{\dot{\theta}}{2}\cos\left(\frac{\theta}{2}\right),$$

$$e_0'' = -\frac{(\theta')^2}{4}\cos\left(\frac{\theta}{2}\right) - \frac{\theta''}{2}\sin\left(\frac{\theta}{2}\right), \ e_3'' = -\frac{(\theta')^2}{4}\sin\left(\frac{\theta}{2}\right) + \frac{\theta''}{2}\cos\left(\frac{\theta}{2}\right), \quad (34)$$

$$\ddot{e}_0 = -\frac{(\dot{\theta})^2}{4}\cos\left(\frac{\theta}{2}\right) - \frac{\ddot{\theta}}{2}\sin\left(\frac{\theta}{2}\right), \ \ddot{e}_3 = -\frac{(\dot{\theta})^2}{4}\sin\left(\frac{\theta}{2}\right) + \frac{\ddot{\theta}}{2}\cos\left(\frac{\theta}{2}\right).$$

Furthermore, the variations of quaternion elements in terms of the variation of rotation angle for the linear formulation can be expressed as

$$\delta e_0 = -\frac{1}{2}\sin\left(\frac{\theta}{2}\right)\delta\theta, \ \delta e_3 = \frac{1}{2}\cos\left(\frac{\theta}{2}\right)\delta\theta. \quad (35)$$

### 2.6.1. Nonlinear formulation

Substituting Eq. (34) into Eqs. (25) and (26), respectively, give

$$-2(\theta')^2\cos\left(\frac{\theta}{2}\right) - 2\theta''\sin\left(\frac{\theta}{2}\right) + \gamma\cos\left(\frac{\theta}{2}\right)(1-s)$$

$$+\left[2\cos\left(\frac{\theta}{2}\right)\int_1^s\int_0^{\hat{s}}\left(-\dot{\theta}^2\cos(\theta) - \ddot{\theta}\sin(\theta)\right)d\bar{s}d\hat{s}\right]$$

$$+\left[2\sin\left(\frac{\theta}{2}\right)\int_1^s\int_0^{\hat{s}}\left(-\dot{\theta}^2\sin(\theta) + \ddot{\theta}\cos(\theta)\right)d\bar{s}ds\right]$$

$$+2U\sqrt{\beta}\left[2\cos\left(\frac{\theta}{2}\right)\int_1^s -\dot{\theta}\sin(\theta)d\hat{s}\right] \quad (36)$$

$$+2U\sqrt{\beta}\left[2\sin\left(\frac{\theta}{2}\right)\int_1^s \dot{\theta}\cos(\theta)d\hat{s}\right]$$

$$-U^2\left[2\cos\left(\frac{\theta}{2}\right)\cos(\theta(1)) + 2\sin\left(\frac{\theta}{2}\right)\sin(\theta(1))\right]$$

$$-\alpha\left(-2\dot{\theta}''\sin\left(\frac{\theta}{2}\right) - 2\dot{\theta}'\theta'\cos\left(\frac{\theta}{2}\right)\right) - 2\cos\left(\frac{\theta}{2}\right)\lambda = 0,$$

$$-2(\theta')^2 \sin\left(\frac{\theta}{2}\right) + 2\theta'' \sin\left(\frac{\theta}{2}\right) - \gamma \sin\left(\frac{\theta}{2}\right)(1-s)$$

$$-\left[2\sin\left(\frac{\theta}{2}\right)\int_1^s \int_0^{\hat{s}} \left(-\dot{\theta}^2 \cos(\theta) - \ddot{\theta}\sin(\theta)\right) d\bar{s} d\hat{s}\right]$$

$$+\left[2\cos\left(\frac{\theta}{2}\right)\int_1^s \int_0^{\hat{s}} \left(-\dot{\theta}^2 \sin(\theta) + \ddot{\theta}\cos(\theta)\right) d\bar{s} d\hat{s}\right]$$

$$-2U\sqrt{\beta}\left[2\sin\left(\frac{\theta}{2}\right)\int_1^s -\dot{\theta}\sin(\theta) d\hat{s}\right] \tag{37}$$

$$+2U\sqrt{\beta}\left[2\cos\left(\frac{\theta}{2}\right)\int_1^s \dot{\theta}\cos(\theta) d\hat{s}\right]$$

$$-U^2\left[-2\sin\left(\frac{\theta}{2}\right)\cos(\theta(1)) + 2\cos\left(\frac{\theta}{2}\right)\sin(\theta(1))\right]$$

$$-\alpha\left(+2\ddot{\theta}''\cos\left(\frac{\theta}{2}\right) - 2\dot{\theta}'\theta'\sin\left(\frac{\theta}{2}\right)\right) - 2\lambda\sin\left(\frac{\theta}{2}\right) = 0.$$

Multiplying Eq. (36) and (37) by, respectively, $(-\delta e_0)$ and $(-\delta e_3)$ given in Eq. (35), and adding the resulting equations, one can obtain the rotation-based governing equation as follows [27, 30, 32]

$$-\theta'' + (1-s)\gamma \sin(\theta)$$

$$+\cos(\theta)\int_1^s \int_0^{\hat{s}} \left((\dot{\theta})^2 \sin(\theta) - \ddot{\theta}\cos(\theta)\right) d\bar{s} d\hat{s}$$

$$-\sin(\theta)\int_1^s \left[\int_0^{\hat{s}} \left((\dot{\theta})^2 \cos(\theta) + \ddot{\theta}\sin(\theta)\right) d\bar{s}\right] d\hat{s} \tag{38}$$

$$-2U\sqrt{\beta}\left(\sin(\theta)\int_1^s \dot{\theta}\sin(\theta) d\hat{s} + \cos(\theta)\int_1^s \dot{\theta}\cos(\theta) d\hat{s}\right)$$

$$-\alpha\ddot{\theta}'' + U^2 \sin(\theta(1) - \theta) = 0.$$

Similarly, one can obtain the following rotation-based boundary conditions from Eq. (28).

$$\theta(0) = 0, \quad \theta'(1) = 0. \tag{39}$$

### 2.6.2. Linear formulation

In view of Eq. (4), for the linear formulation, $e_3$ can be approximated by $\theta/2$. Recalling the approximated value of $e_0 \approx 1$ for the linear formulation, the variation of quaternion elements in Eq. (35) can be approximated by

$$\delta e_0 \approx -\frac{\theta}{4}\delta\theta, \quad \delta e_3 \approx \frac{1}{2}\delta\theta. \tag{40}$$

Therefore, Eq. (32) can be rewritten as

$$2\theta'' - \gamma(1-s)\theta + \left[2\int_1^s \int_0^{\hat{s}} \ddot{\theta}\,d\tilde{s}\,d\hat{s}\right] + 2U\sqrt{\beta}\left[2\int_1^s \dot{\theta}\,d\hat{s}\right] \\ -U^2[-\theta + 2\theta(1)] - 2\alpha\dot{\theta}'' - \lambda\theta = 0. \tag{41}$$

Multiplying Eq. (31) and (41) by, respectively, $(-\delta e_0)$ and $(-\delta e_3)$ given in Eq. (40), and adding the resulting equations gives the linear form of the rotation-based governing equation [27, 32]

$$-\theta'' + \gamma(1-s)\theta - \int_1^s \int_0^{\hat{s}} \ddot{\theta}\,d\tilde{s}\,d\hat{s} - 2U\sqrt{\beta}\int_1^s \dot{\theta}\,d\hat{s} - \alpha\dot{\theta}'' + U^2\left(\theta(1) - \theta\right) = 0. \tag{42}$$

Similar to the quaternion-based mathematical model, the boundary conditions for the rotation-based linear formulation are similar to the nonlinear ones given in Eq. (39).

## 2.7. Period of oscillation

Chen et al. [27] mathematically proved that for the self-excited oscillations of pipe conveying fluid in the post-flutter region based on the geometrically exact rotation-based model, the oscillation period of $\theta$, $T_\theta$, is equal to the oscillation period of $\eta$, $T_\eta$, and they are twice the oscillation period of $\zeta$, $T_\zeta$, i.e., $T_\theta = T_\eta = 2T_\zeta$. For the geometrically exact quaternion-based model, in view of Eq. (4), it can easily be proved that the oscillation period of $e_3$, $T_{e_3}$, is equal to $T_\theta$, and they are twice the oscillation period of $e_0$, $T_{e_0}$, i.e., $T_\theta = T_{e_3} = 2T_{e_0}$. Similar to Chen et al. [43], in light of Eq. (30), one can obtain $T_\theta = T_{e_3} = T_\eta = 2T_{e_0} = 2T_\zeta$.

## 3. Method of solution

The index of integro-partial differential-algebraic equations listed in Eqs. (24)-(26) is three (see Ref. [43] for the definition of the index). In order to obtain a set of ordinary differential-algebraic equations with index one from the partial differential-algebraic equations, the constraint equation is differentiated twice with respect to $\tau$.

$$\dot{e}_0^2 + \dot{e}_3^2 + e_0\ddot{e}_0 + e_3\ddot{e}_3 = 0. \tag{43}$$

Therefore, Eq. (43) as the new constraint equation is employed rather than the original constraint equation given in Eq. (26). It should be pointed out that the numerical solution of the new system needs no longer to satisfy the original constraint equation precisely [43]. Hence, the original constraint equation should be checked after the numerical solution.

$$C = e_0^2 + e_3^2 - 1 \tag{44}$$

In order to convert the partial differential algebraic equations given in Eqs. (25), (26), and (43) into a set of ordinary differential algebraic equations based on the Galerkin discretization technique, the elements of the quaternion system and the Lagrangian multiplier function, respectively, are approximated by

$$e_0 = 1 + \sum_{n=1}^{N_0} \phi_n p_n, \quad e_3 = \sum_{n=1}^{N_3} \psi_n q_n, \quad \lambda = \sum_{n=1}^{N_\lambda} \chi_n r_n \tag{45}$$

in which $\phi_n$, $\psi_n$, and $\chi_n$ denote the $n$th approximation function for $e_0$, $e_3$, and $\lambda$, respectively; $p_n$, $q_n$, and $r_n$ represent the $n$th generalized coordinate of $e_0$, $e_3$, and $\lambda$, respectively. Additionally, $N_0$, $N_3$, and $N_\lambda$ stand for the number of approximation functions for $e_0$, $e_3$, and $\lambda$, respectively.

In this study, the following approximation functions are utilized

$$\phi_n = \sin\left(\frac{2n-1}{2}\pi s\right), \psi_n = \sin\left(\frac{2n-1}{2}\pi s\right), \chi_n = \sin\left(\frac{n\pi}{4}\left(s + \frac{1}{2}\right)\right). \tag{46}$$

It is evident that the boundary conditions listed in Eq. (29) for $e_0$ and $e_3$ are satisfied by the model utilized for the discretization. Moreover, it should be pointed out that the Lagrangian multiplier function, $\lambda$, is boundary conditions-free. Finally, it should be noticed that in this study, it is assumed that $N_0 = N_3 = N_\lambda = N$.

### 3.1. Nonlinear model for dynamic analysis

Substituting Eq. (45) into Eqs. (24), (25), and (43), respectively, give

$$\begin{aligned}
&K^{11}_{ijkl} p_j q_k q_l + {}^1K^{11}_{ijk} q_j q_k + K^{12}_{ij} p_j + {}^1K^{12}_i + M^{11}_{ijkl} p_j p_k \ddot{p}_l + {}^1M^{11}_{ijk} p_j \ddot{p}_k + {}^2M^{11}_{ij} \ddot{p}_j + \\
&M^{12}_{ijkl} p_j \dot{p}_k \dot{p}_l + {}^1M^{12}_{ijk} \dot{p}_j \dot{p}_k + M^{13}_{ijkl} p_j q_k \ddot{q}_l + {}^1M^{13}_{ijk} q_j \ddot{q}_k + M^{14}_{ijkl} p_j \dot{q}_k \dot{q}_l + {}^1M^{14}_{ijk} \dot{q}_j \dot{q}_k + \\
&M^{15}_{ijkl} q_j p_k \ddot{q}_l + {}^1M^{15}_{ijk} q_j \ddot{q}_k + M^{16}_{ijkl} q_j \dot{p}_k \dot{q}_l + M^{17}_{ijkl} q_j q_k \ddot{p}_l + G^{11}_{ijkl} p_j p_k \dot{p}_l + {}^1G^{11}_{ijk} p_j \dot{p}_k + \\
&{}^2G^{11}_{ij} \dot{p}_j + G^{12}_{ijkl} p_j q_k \dot{q}_l + {}^1G^{12}_{ijk} q_j \dot{q}_k + G^{13}_{ijkl} q_j q_k \dot{p}_l + G^{14}_{ijkl} q_j \dot{q}_k p_l + {}^1G^{14}_{ijk} q_j \dot{q}_k + \\
&C^{11}_{ijkl} q_j q_k \dot{p}_l + C^{12}_{ijkl} q_j p_k \dot{q}_l + C^1_{12} q_j \dot{q}_k + K^{13}_{ijkl} p_j p_k p_l + {}^1K^{13}_{ijk} p_j p_k + {}^2K^{13}_{ij} p_j + {}^3K^{13}_i + \\
&K^{14}_{ijkl} p_j q_k q_l + {}^1K^{14}_{ijk} q_j q_k + K^{15}_{ijkl} q_j p_k q_l + {}^1K^{15}_{ijk} q_j q_k + K^{16}_{ijk} p_j r_k + {}^1K^{16}_{ij} r_j = 0, \quad i,j,k,l = 1,2,..,N;
\end{aligned} \tag{47}$$

$$\begin{aligned}
&K^{21}_{ijkl} p_j p_k q_l + {}^1K^{21}_{ijk} p_j q_k + \left({}^2K^{21}_{ij} + K^{22}_{ij}\right) q_j + M^{21}_{ijkl} q_j p_k \ddot{p}_l + {}^1M^{21}_{ijk} q_j \ddot{p}_k + M^{22}_{ijkl} q_j \dot{p}_k \dot{p}_l + \\
&M^{23}_{ijkl} q_j q_k \ddot{q}_l + M^{24}_{ijkl} q_j \dot{q}_k \dot{q}_l + M^{25}_{ijkl} p_j p_k \ddot{q}_l + {}^1M^{25}_{ijk} p_j \ddot{q}_k + {}^2M^{25}_{ij} \ddot{q}_j + M^{26}_{ijkl} p_j \dot{p}_k \dot{q}_l + \\
&{}^1M^{26}_{ijk} \dot{p}_j \dot{q}_k + M^{27}_{ijkl} p_j q_k \ddot{p}_l + {}^1M^{27}_{ijk} q_j \ddot{p}_k + G^{21}_{ijkl} q_j p_k \dot{p}_l + {}^1G^{22}_{ijk} q_j \dot{p}_k + G^{22}_{ijkl} q_j q_k \dot{q}_l + \\
&G^{23}_{ijkl} p_j q_k \dot{p}_l + {}^1G^{23}_{ijk} q_j \dot{p}_k + G^{24}_{ijkl} p_j \dot{q}_k p_l + {}^1G^{24}_{ijk} p_j \dot{q}_k + {}^2G^{24}_{ij} \dot{q}_j + C^{21}_{ijkl} p_j q_k \dot{p}_l + \\
&{}^1C^{21}_{ijk} q_k \dot{p}_l + C^{22}_{ijkl} p_j p_k \dot{q}_l + {}^1C^{22}_{ijk} p_j \dot{q}_k + {}^2C^{22}_{ij} \dot{q}_j + K^{23}_{ijk} q_j p_k p_l + {}^1K^{23}_{ijk} q_j p_k + \\
&{}^2K^{23}_{ij} q_j + K^{24}_{ijkl} q_j q_k q_l + K^{25}_{ijkl} p_j p_k q_l + {}^1K^{25}_{ijk} p_j q_k + {}^2K^{25}_{ij} q_j + K^{26}_{ijk} q_j r_k = 0, \quad i,j,k,l = 1,2,..,N;
\end{aligned} \tag{48}$$

$$D^{31}_{ijk} \dot{p}_j \dot{p}_k + D^{32}_{ijk} \dot{q}_j \dot{q}_k + D^{33}_{ijk} p_j \ddot{p}_k + {}^1D^{33}_{ij} \ddot{p}_k + D^{34}_{ijk} q_j \ddot{q}_k = 0, \quad i,j,k = 1,2,..,N. \tag{49}$$

in which the coefficients listed in Eqs. (47)-(49) are defined in [Appendix A](#).

In order to numerically determine the nonlinear geometrically exact dynamics of the system, the ordinary differential algebraic equations given in Eqs. (47)-(49) were solved by using the ode15s solver in MATLAB software.

## 3.2. Linear model for stability analysis

Substituting Eq. (45) into Eq. (41) leads to

$$[M]\{\ddot{q}\}+[C]\{\dot{q}\}+[K]\{q\}=0, \tag{50}$$

where

$$[M]=\left[{}^2M^{25}\right],\ [C]=\left[{}^2G^{24}\right]+\left[{}^2C^{22}\right],\ [K]=\left[K^{12}\right]+2\left[K^{22}\right]+2\left[{}^2K^{23}\right]+\left[{}^2K^{25}\right]. \tag{51}$$

The solution of Eq. (50) for determining the stability characteristics of system can be expressed as

$$\{q\}=\{q_0\}e^{\omega\tau}, \tag{52}$$

where $\omega$ is a complex non-dimensional eigenvalue and $q_0$ is the oscillation amplitude. Substituting Eq. (52) into Eq. (50) results in

$$\left([M]\omega^2+[C]\omega+[K]\right)\{\lambda\}=0. \tag{53}$$

The quadratic eigenvalue problem given in Eq. (53) can be reformulated as a linear one as follows

$$[A]\{y\}=\omega\{y\}, \tag{54}$$

in which

$$[A]=\begin{bmatrix}[0] & [I]\\ -[M]^{-1}[K] & -[M]^{-1}[C]\end{bmatrix},\ \{y\}=\begin{Bmatrix}\{q_0\}\\ \omega\{q_0\}\end{Bmatrix}. \tag{55}$$

The natural frequencies and the critical flow velocity at which the system becomes unstable via a flutter instability can be determined via the real and imaginary parts of $\omega$.

## 4. Results and discussion

It is acknowledged that a hanging cantilevered pipe conveying fluid, which is described in Section 2.1, loses its stability at a critical flow velocity via flutter instability, where the critical flow velocity in the dimensionless form, $U_{cr}$, is dependent on the mass parameter, $\beta$, gravity parameter, $\gamma$, and the dimensionless damping coefficient, $\alpha$, [3, 21]. If the system in the pre-flutter region, $U < U_{cr}$, undergoes an initial excitation, after a while the amplitude of oscillation becomes zero; no matter how large the initial excitation [3, 21, 27, 30]. For the post-flutter region, $U > U_{cr}$, a vise-versa scenario occurs; no matter how small the initial excitation, the system undergoes a self-excited oscillation, and the dimensionless flow velocity, mass parameter, gravity parameter, and dimensionless damping coefficient influence the oscillation amplitude and period [3, 21, 27, 30]. Therefore, this section compares the critical flow velocity and post-flutter dynamics of hanging cantilevered flexible pipe conveying fluid using the geometrically exact quaternion-based model developed in this study with those reported via the geometrically exact rotation-based model. Lastly, the advantages and challenging issues of the newly developed model are discussed.

Prior to conducting the comparative studies, it should be mentioned that for the nonlinear system with $N$ modes, $2N$ first-order ordinary differential equations should be solved for the geometrically exact rotation-based model, but for the geometrically exact quaternion-based model, $5N$ first-order ordinary differential algebraic equations should be solved based on the solution strategy chosen in this study.

The dimensionless critical flow velocities, $U_{cr}$, as a function of the mass parameter, $\beta$, obtained by the current model are compared with those reported by Chen et al. [27] and Dehrouyeh-Semnani [32] based on the linear rotation-based model, i.e., Eq. (42) and the Galerkin technique. The plots in Figure 2a illustrate that the results of the present model are in excellent agreement with those obtained by Chen et al. [27]. The reason for such a fantastic agreement for each arbitrary value of $N$ is that Chen et al. [27] utilized $\psi_n$ to approximate $\theta$ in Eq. (42). Figure 2b indicates that the dimensionless critical velocities predicted in this study for a system with structural damping can verify those achieved by Dehrouyeh-Semnani [32]. It should be pointed out that Dehrouyeh-Semnani [32] used the first derivatives of linear mode shapes of a cantilevered Euler-Bernoulli beam as the approximation functions. Further comparative studies for the dimensionless critical flow velocity are presented in Figure 3. The plots in the figure show that the current model is capable of verifying the results reported by Ref. [3] based on the linear model derived from the transverse motion.

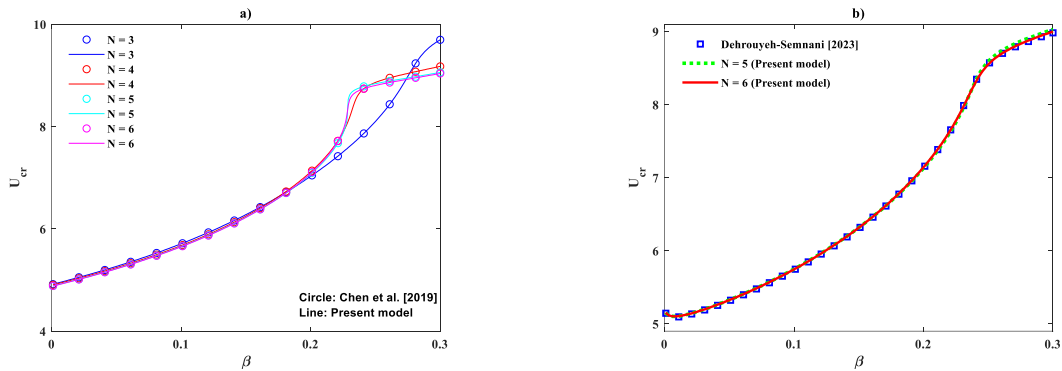

Figure 2: Comparison of dimensionless critical flow velocities for $\gamma = 18.9$, $0 < \beta \leq 0.3$ and a) $\alpha = 0$, b) $\alpha = 0.0018$.

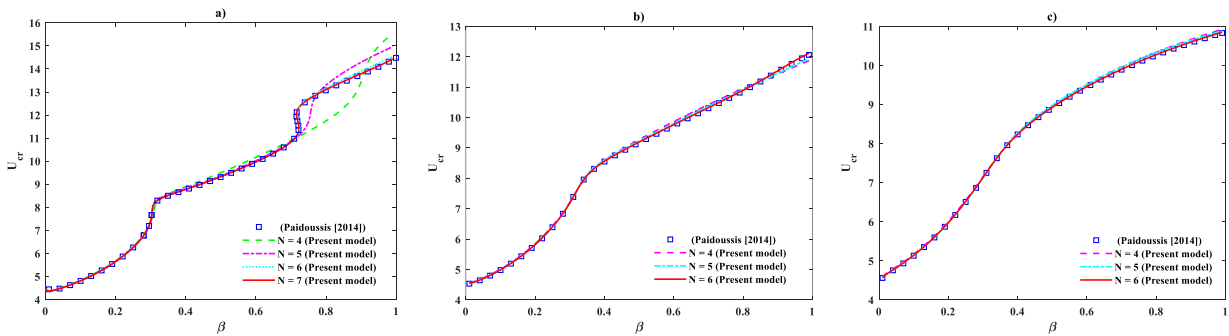

Figure 3: Comparison of dimensionless critical flow velocities for γ = 0, 0 < β < 1, and a) α = 0.001, b) α = 0.005, c) α = 0.01.

Figure 4 is constructed to indicate the nonlinear response of system to a large initial condition when the flow velocity is less than the critical flow velocity ($U < U_{cr}$). According to the linear stability analysis, the system is stable at such flow velocities; therefore, the response of system tends to zero. The plots depicted in the figure display that the nonlinear behavior of system obeys the prediction based on the linear model. Additionally, it can be seen that for the system with a flow velocity closer to the critical one, more dimensionless time is needed to reach its final state.

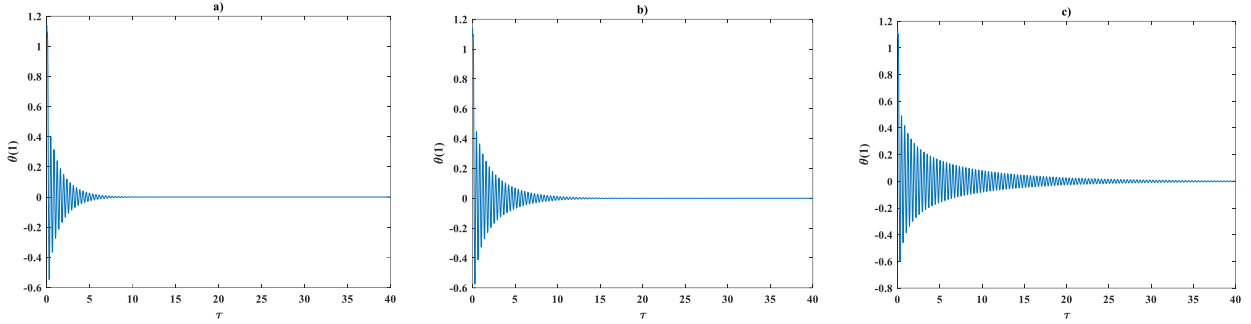

Figure 4: Time traces of $\theta(1)$ when $\beta = 0.142$, $\gamma = 18.9$, $\alpha = 0$, and $U < U_{cr}$; a) $U = 5.9$, b) $U = 6.0$, and c) $U = 6.1$.

The first verification study for the nonlinear geometrically exact dynamics of cantilevered pipe conveying fluid is devoted to the time history of pipe tip when $\beta = 0.142$, $\gamma = 18.9$, $\alpha = 0$, and $U = 6.5$. The plots depicted in Figure 5 show that the time histories of tip slope, $\theta(1)$, tip dimensionless transverse displacement, $\eta(1)$, and tip dimensionless longitudinal displacement, $\zeta(1)$, based on the geometrically exact quaternion-based model are in very good agreement with $\theta(1)$ and $\zeta(1)$ reported by Chen et al. [27] (see Figs. 7a and 8a in their work) according to the geometrically exact rotation-based model, i.e., Eq. (38). These time histories indicate the self-excited behavior of system in the post-flutter region in which a very small initial excitation leads to a steady-state harmonic motion with large deformations. Furthermore, the time histories of the quaternion elements at the pipe tip are plotted in Figure 6a-b. Figure 6c investigates the original constraint equation at the pipe tip based on Eq. (44). It can be seen that the constraint is satisfied very well.

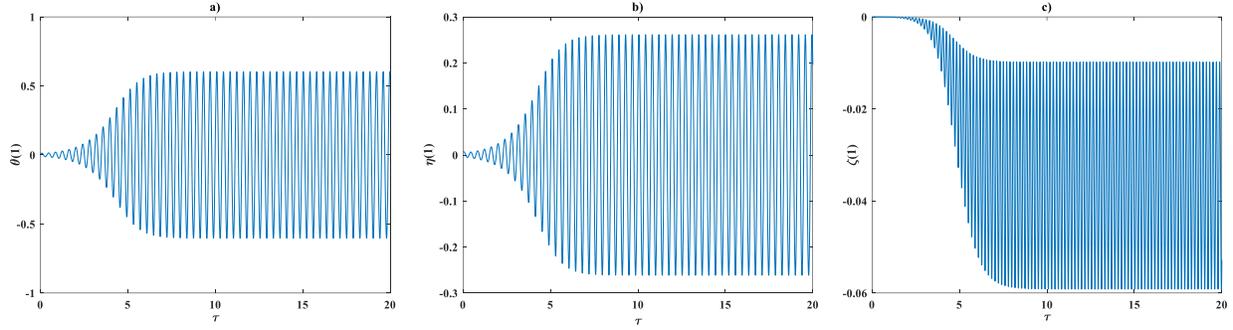

Figure 5: Time traces of $\theta(1)$, $\eta(1)$, and $\zeta(1)$ when $\beta = 0.142$, $\gamma = 18.9$, $\alpha = 0$, and $U = 6.5$.

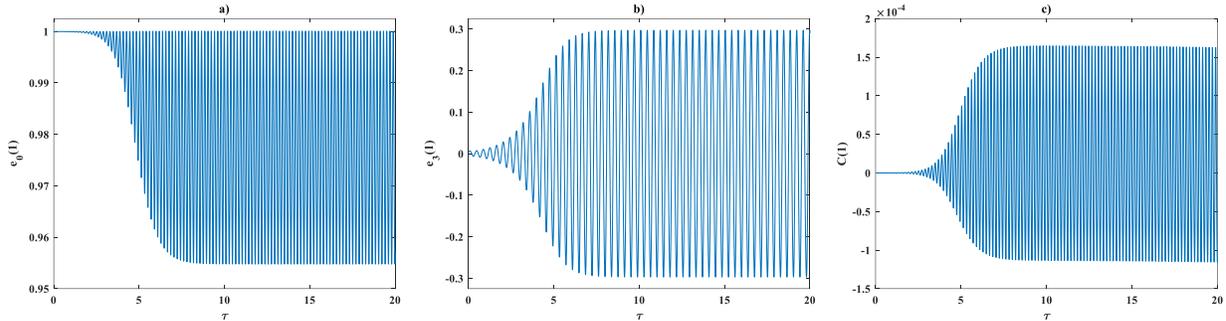

Figure 6: Time trace for $e_0(1)$ and $e_3(1)$, along with $C(1)$ corresponding to Figure 5.

Figure 7 displays the nonlinear geometrically exact dynamic response of the cantilevered pipe conveying fluid in the dimensionless flow velocity range of [6 10] for $\beta = 0.142$, $\gamma = 18.9$, and $\alpha = 0$. The bifurcation diagrams plotted in the figure indicate the tip slope, $\theta(1)$, tip dimensionless longitudinal displacement, $\zeta(1)$, and tip dimensionless transverse displacements, $\eta(1)$, based on the present model with $N = 3, 4,$ and 5. It should be pointed out that the dimensionless critical flow velocities for $N = 3, 4$, and 5, are as follows $U_{cr} = 6.175, 6.1425,$ and $6.1225$, respectively. Therefore, the starting points of the bifurcation disarms in Figure 7 are slightly different. The plotted results in the figure show that for $N = 4$, the bifurcation diagrams of $\theta(1)$ and $\zeta(1)$ predicted via the current model verifies very well those reported by Chen et al. [27] (see Figs. 5a-b and 6 in their work for both $N = 3$ and 4) based on the geometrically exact rotation-based model with $N = 4$. Additionally, in the case of $N = 3$, the results of both models are in good agreement except for the dimensionless flow velocities close to $U = 10$ where the results differ slightly. As the figure shows, the bifurcation diagrams plotted based on the present model with $N = 5$ verify the obtained results with $N = 4$ except for the dimensionless flow velocities close to $U = 10$, but the differences are relatively slight. Moreover, the corresponding bifurcation diagrams of the quaternion elements are depicted in Figure 8a-b. Maximum/minimum values of the original constraint equation at the pipe tip, $C(1)$, for the system under investigation are depicted in Figure 8c. The plots in the figure show that for $N = 4$ and 5, $C(1)$ is in the range of [-0.005 0.005], and the obtained results for these mode numbers are acceptable. However, for $N = 3$, when $U$ is about 8, maximum $C(1)$ approaches 0.005, and for $U = 10$, the minimum and maximum of $C(1)$ is about 0.01 and 0.025, respectively. Therefore, the results of $N = 3$ are less accurate than those obtained via $N = 4$ and 5.

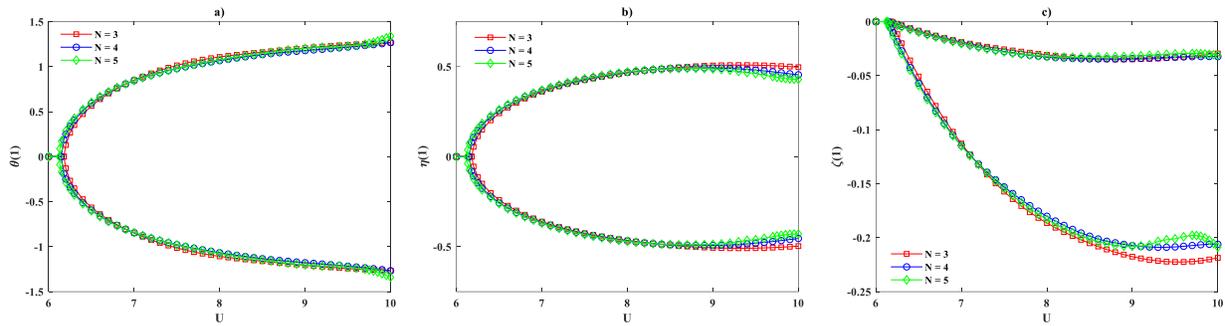

Figure 7: Maximum/minimum of $\theta(1)$, $\eta(1)$, and $\zeta(1)$ when $\beta = 0.142$, $\gamma = 18.9$, and $\alpha = 0$.

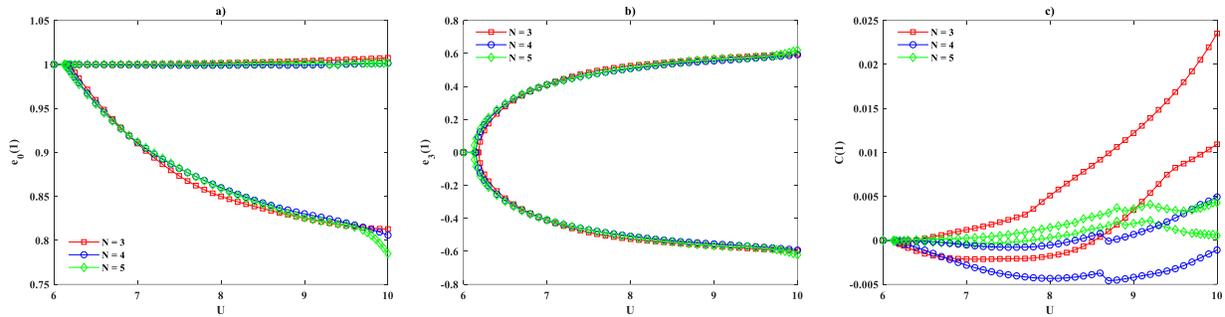

Figure 8: Maximum/minimum of $e_0(1)$, $e_3(1)$, and $C(1)$ corresponding to Figure 7.

Figure 9 is constructed to conduct a comparison study for the phase planes of $\theta(1)$, $\eta(1)$, and $\zeta(1)$ for the system with $\beta = 0.142$, $\gamma = 18.9$, $\alpha = 0$, and different values of the dimensionless flow velocities $U = 6.3, 6.5, 7.0$, and $8.5$. Again, the achieved results based on the geometrically exact rotation-based model with $N = 4$ by Chen et al. [27] ( see Figs 7b, 8b, and 9c in their work) are utilized to verify the results of the present model with $N = 4$ and 5. The plots depicted in Figure 9 illustrate that the results of present model with $N = 4$ and 5 are in good agreement. Moreover, the current model verifies the phase planes obtained by Chen et al. [27]. Furthermore, the corresponding phase planes of the quaternion elements for $N = 4$ and 5 are plotted in Figure 10.

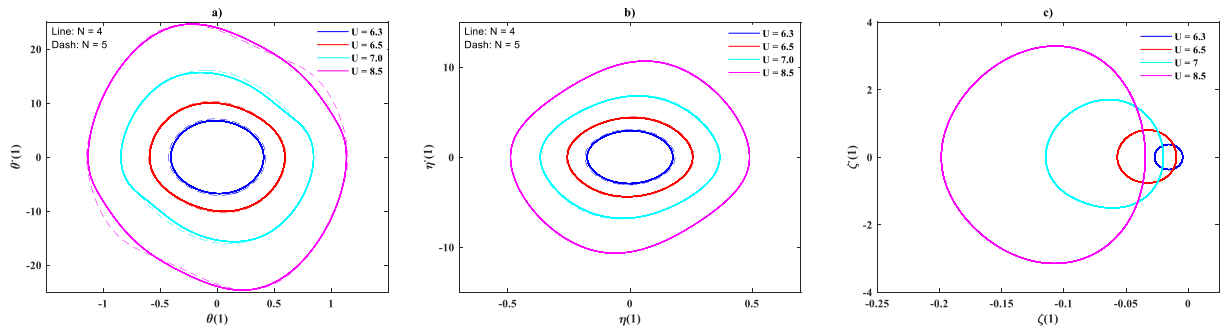

Figure 9: Phase planes of $\theta(1)$, $\eta(1)$, and $\zeta(1)$ for different dimensionless flow velocities corresponding to Figure 7.

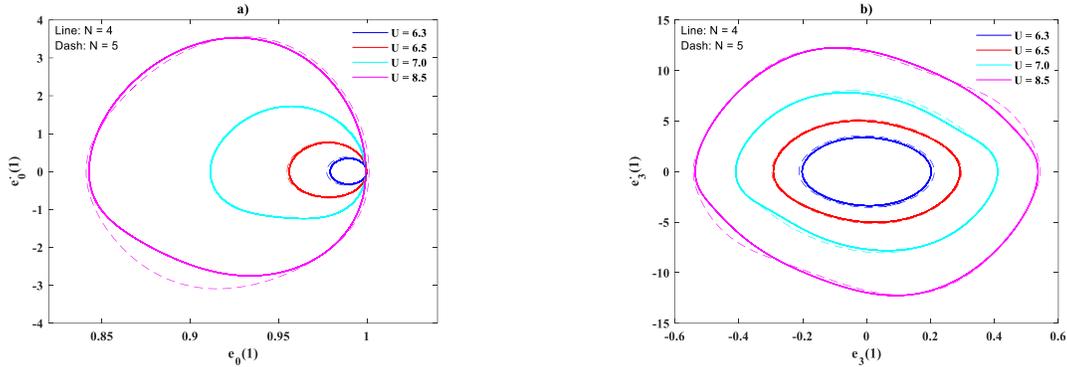

Figure 10: Phase planes of $e_0(1)$ and $e_3(1)$ corresponding to Figure 9.

Figures 11, 12, and 13 indicate that the time histories of system with $U = 6.3, 7.0,$ and $8.5$ when $\beta = 0.142$, $\gamma = 18.9$, and $\alpha = 0$ for $\theta(1)$, $\zeta(1)$, and $\eta(1)$, respectively. These plots are corresponded to the phase planes depicted in Figure 9. The results illustrate that for the system with a flow velocity beyond the critical one, the transient response lasts at a lower dimensionless time when the flow velocity is larger.

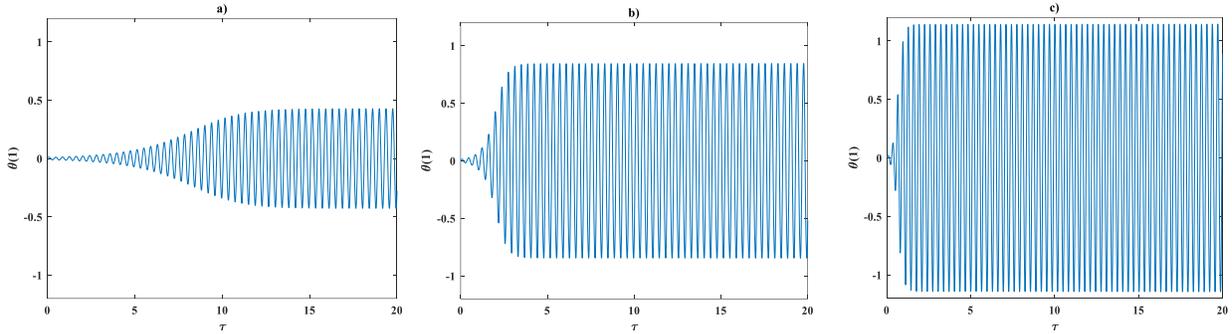

Figure 11: Time traces of $\theta(1)$ when $\beta = 0.142$, $\gamma = 18.9$, $\alpha = 0$, and a) $U = 6.3$, b) $U = 7.0$, c) $U = 8.5$.

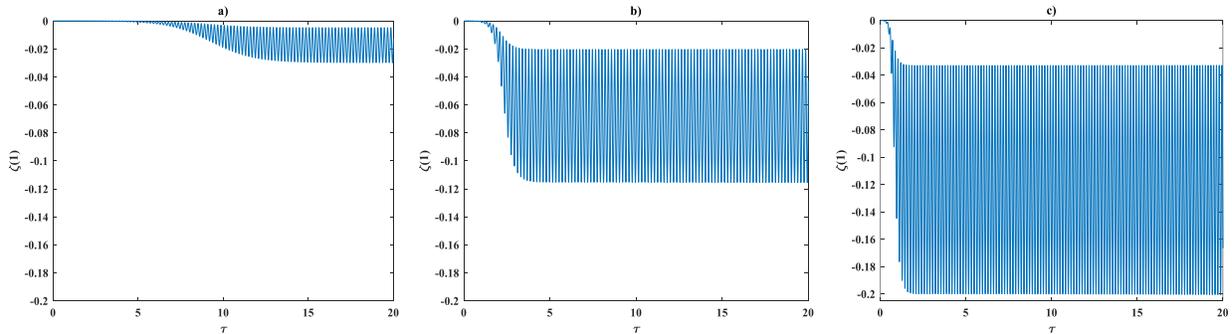

Figure 12: Time traces of $\zeta(1)$ when $\beta = 0.142$, $\gamma = 18.9$, $\alpha = 0$, and a) U = 6.3, b) U = 7.0, c) U = 8.5.

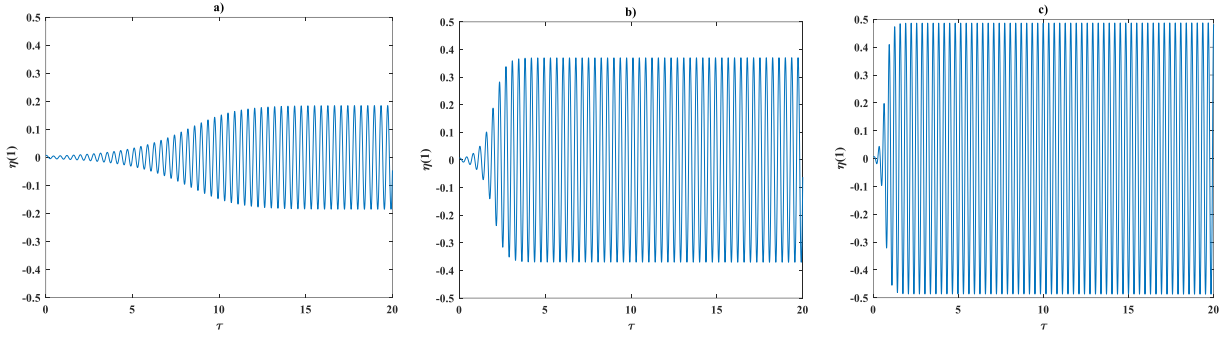

Figure 13: Time traces of $\eta(1)$ when $\beta = 0.142$, $\gamma = 18.9$, $\alpha = 0$, and a) U = 6.3, b) U = 7.0, c) U = 8.5.

For the system under investigation in Figure 7, it is discussed that the present results with $N = 4$ and 5 are slightly different when the dimensionless flow velocity is close to 10. Therefore, the geometrically exact dynamics of system with $U = 10$ are studied in Figure 14-11. The time history of $\eta(1)$ for $\Delta\tau = \tau - \tau_0 = 2$ of the steady-state oscillation are plotted in Figure 14. The figure contains the results of present model for $N = 3, 4$ and 5. It can be seen that the oscillation amplitude of the present model with $N = 4$ is similar to that reported by Chen et al. [27] based on $N = 4$ (see Fig. 10a in their work) and their oscillation periods are slightly different, but very close. The plots also show that the oscillation period of the present model with $N = 3$ is similar to that of Chen et al. [27] based on $N = 4$, but their oscillation amplitudes are slightly different. Additionally, comparing the results of present model with $N = 4$ and 5 indicate that their oscillation amplitudes and periods differ slightly. For further investigation, the deformed shapes of pipe over a cycle of steady-state oscillation based on the present model are shown in Figure 15. The depicted shapes show that the results of the current model with $N = 4$ verify those of Chen et al. [27] based on $N = 4$ (see Fig. 10b in their work). However, the deformed shapes depicted according to the current model with $N = 5$ are slightly different from those of $N = 4$.

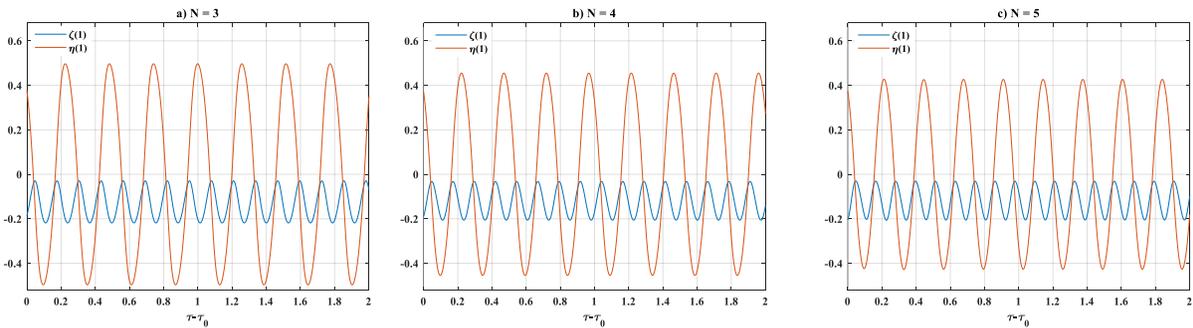

Figure 14: Time traces of $\eta(1)$ and $\zeta(1)$ for different values of $N$ when $\beta = 0.142$, $\gamma = 18.9$, $U = 10$, and $\alpha = 0$.

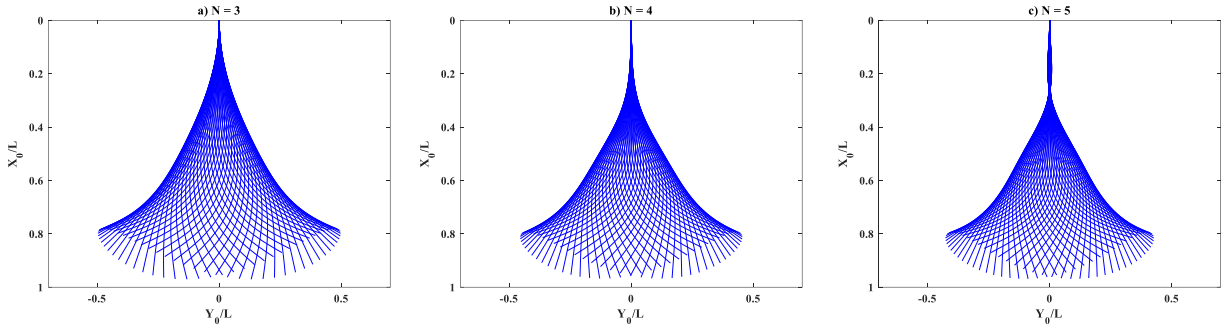

Figure 15: Pipe's deformed configurations over a cycle of steady-state oscillation corresponding to Figure 14.

Additionally, the corresponding time histories of the quaternion elements for $N = 3, 4,$ and 5, along with the corresponding phase planes for $N = 3, 4,$ and 5 are shown in Figure 16 and Figure 17, respectively.

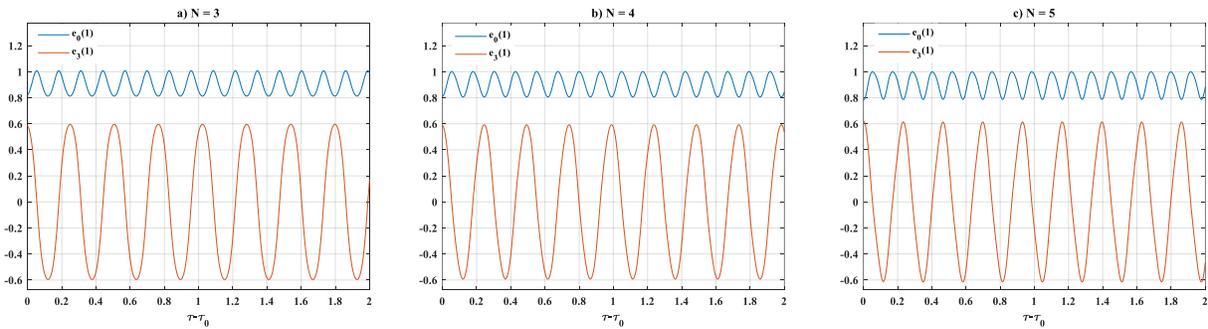

Figure 16: Time traces of $e_0(1)$ and $e_3(1)$ corresponding to Figure 14.

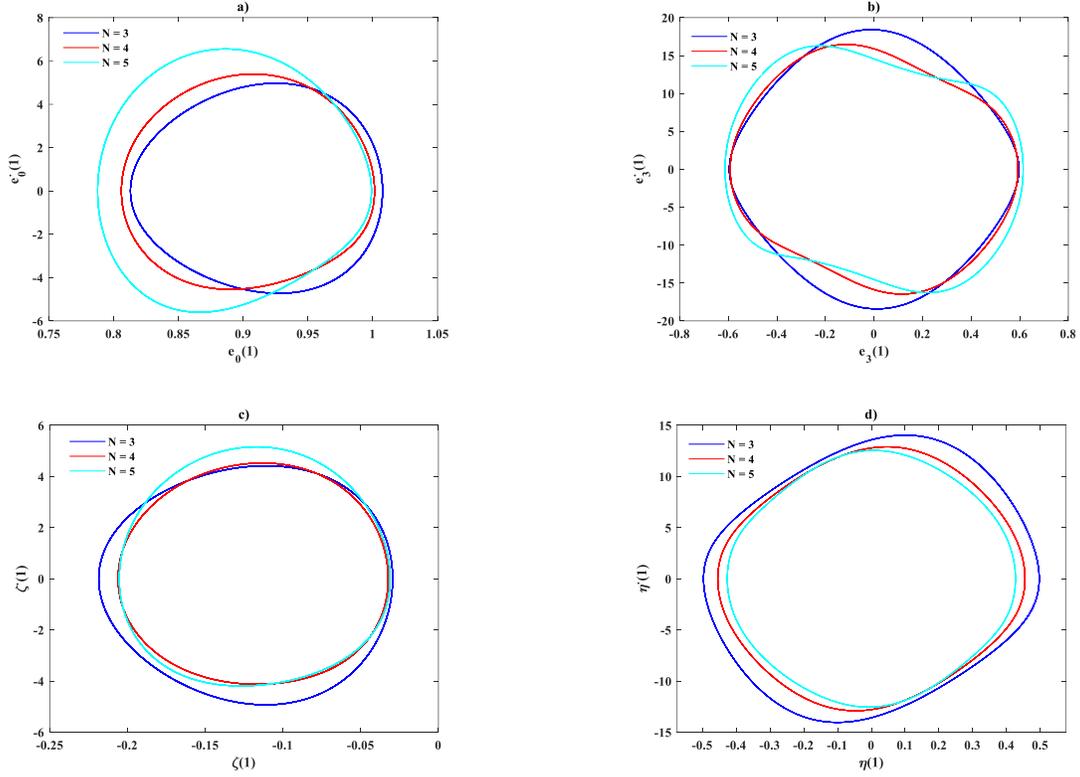

Figure 17: Phase planes of the pipe tip for different dimensionless flow velocities corresponding to Figure 14.

As the last verification study, the geometrically exact rotation-based deformed shapes of cantilevered pipe conveying fluid over a cycle of steady-state oscillation for $\beta = 0.142$, $\gamma = 18.9$, $\alpha = 0.0018$, and $U = 6.6, 7.65, 8.94,$ and $11.1$ proposed by Farokhi et al. [30] are utilized to verify the current model with $N = 5$. The reported results by Farokhi et al. [30] are based on the Galerkin discretization method with $N = 10$ and the first derivatives of linear mode shapes of a cantilevered Euler-Bernoulli beam as the approximation functions. The plots depicted in Figure 18 show that the geometrically exact deformed shapes of the system according to the quaternion-based model are in good agreement with those obtained by Farokhi et al. [30] (see Fig. 6a-d in their work) via the rotation-based model. Additionally, the phase planes of the system for $e_0(1)$, $e_3(1)$, $\zeta(1)$, and $\eta(1)$ are shown in Figure 19. The corresponding time history of the system over a cycle of steady-state oscillation are also plotted in Figure 20. Furthermore, the time histories of $\zeta(1)$ and $\eta(1)$ are demonstrated in Figure 21 and Figure 22, respectively.

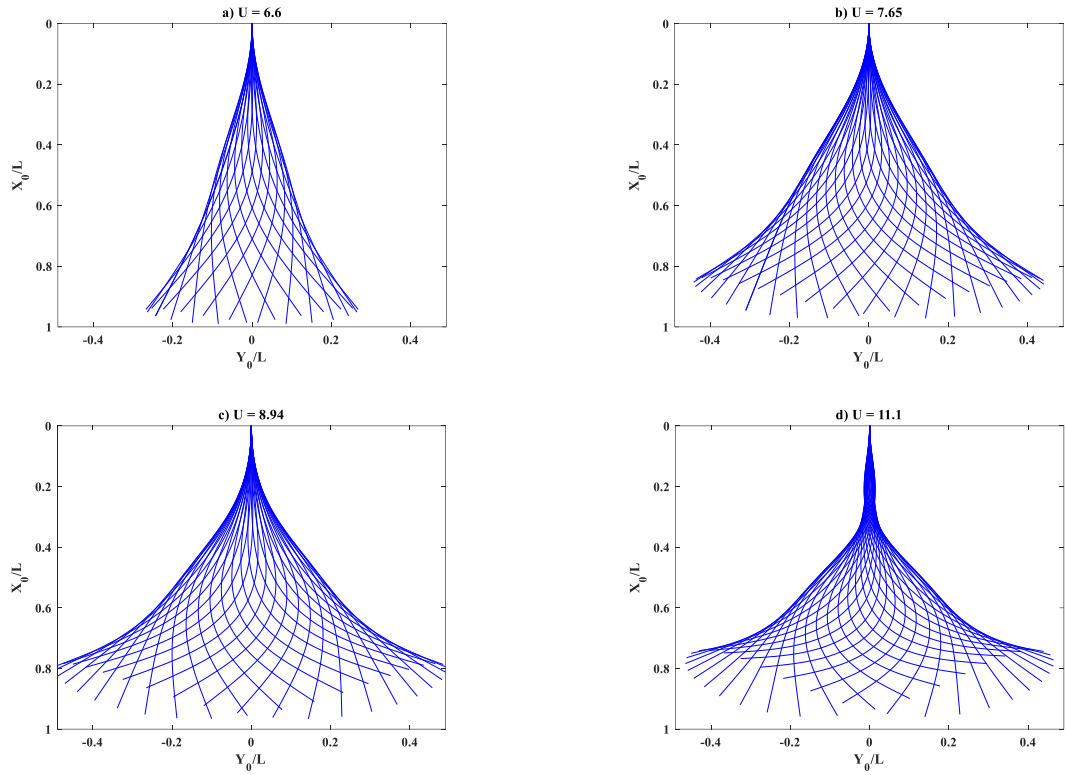

Figure 18: Pipe's deformed configuration over a cycle of steady-state oscillation for different values of dimensionless flow velocities when $\beta = 0.142$, $\gamma = 18.9$, and $\alpha = 0.0018$.

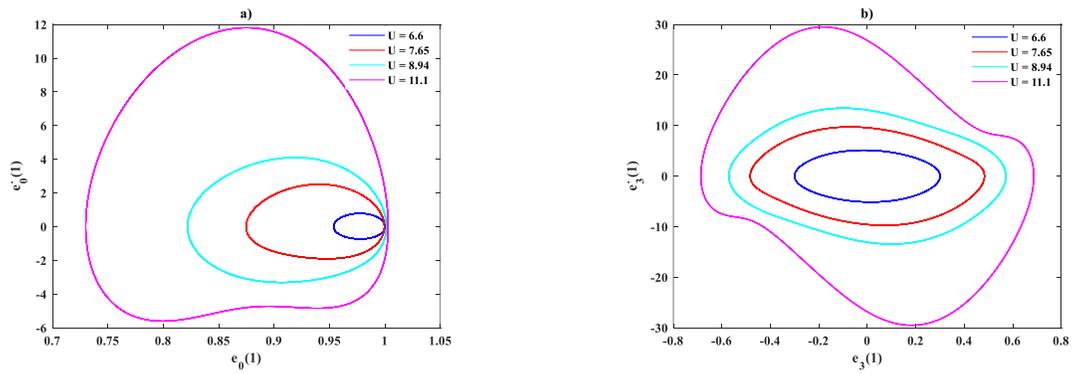

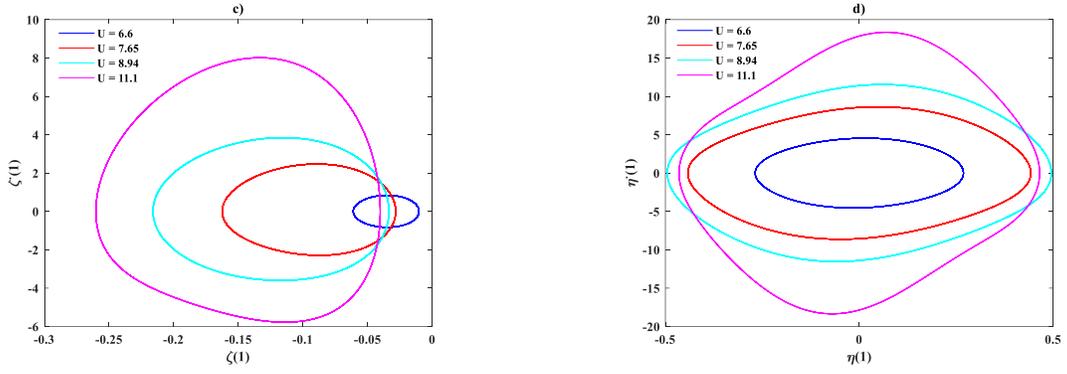

Figure 19: Phase planes of the pipe tip for different dimensionless velocities corresponding to Figure 18.

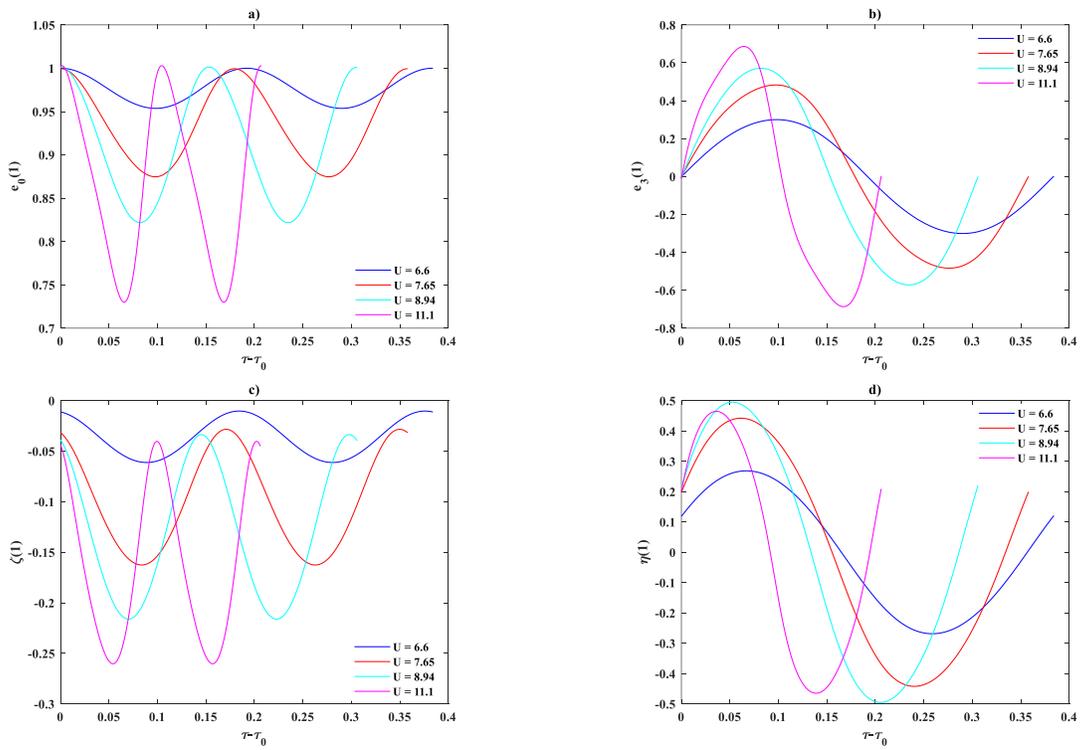

Figure 20: Time traces of the pipe tip over a cycle of steady-state oscillation for different dimensionless velocities corresponding to Figure 18.

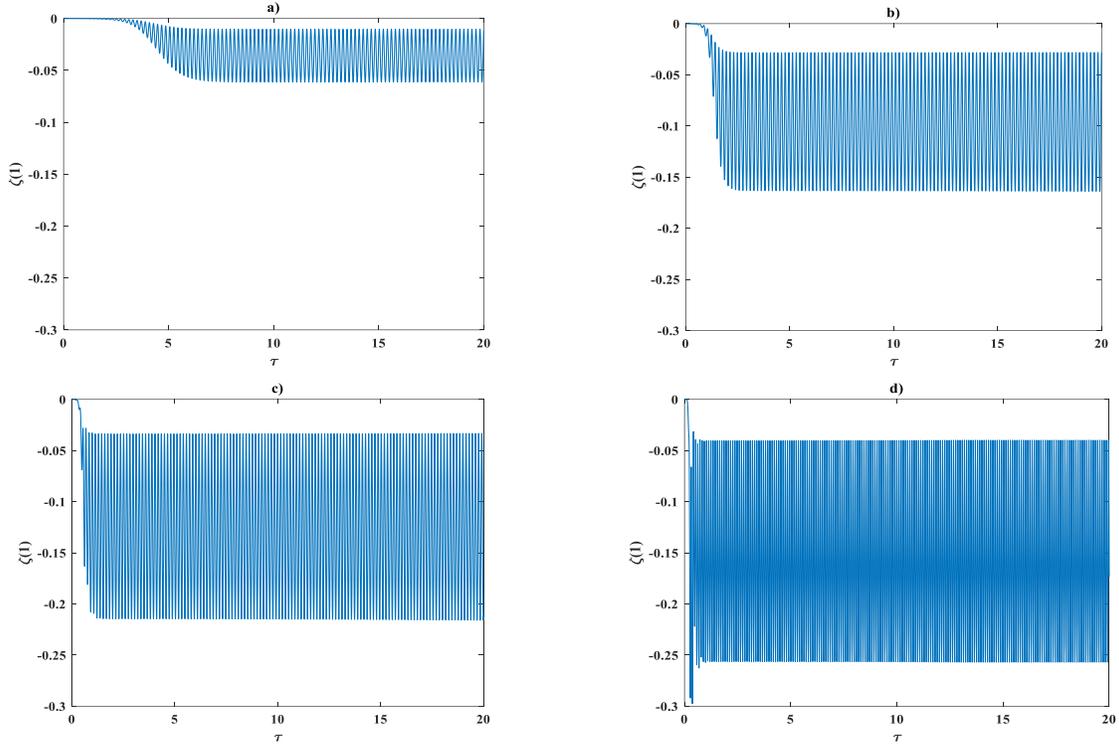

Figure 21: Time traces of $\zeta(1)$ when $\beta = 0.142$, $\gamma = 18.9$, $\alpha = 0.0018$, and a) $U = 6.6$, b) $U = 7.65$, c) $U = 8.94$, d) $U = 11.1$.

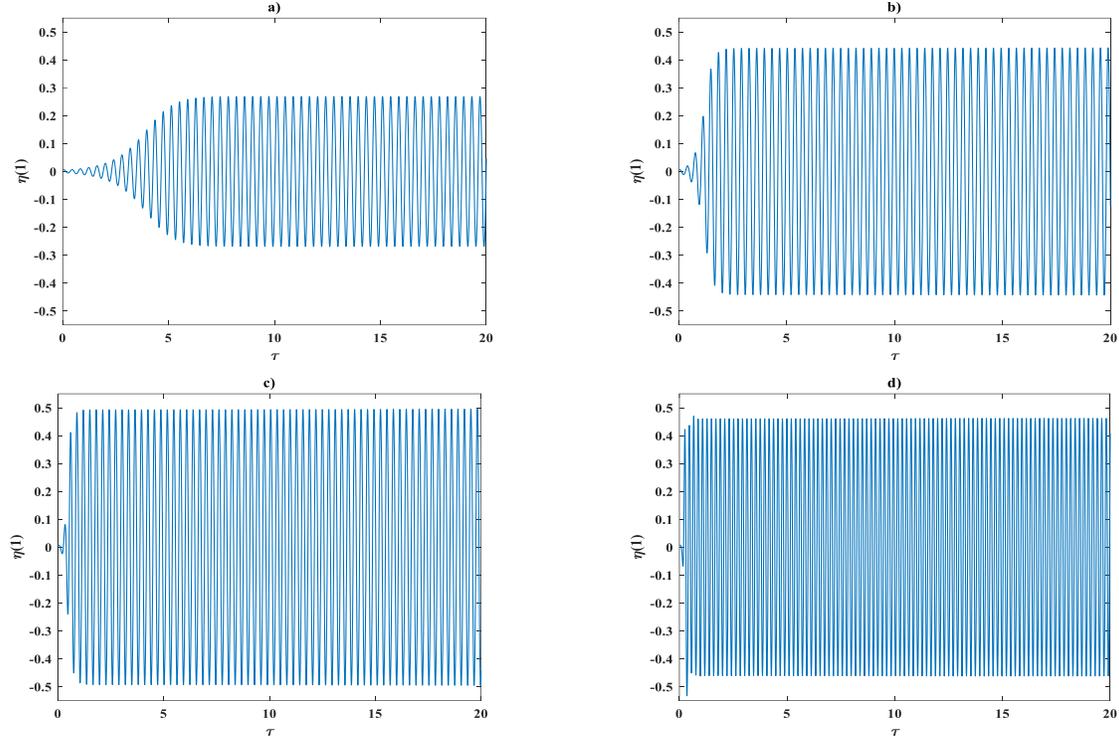

Figure 22: Time traces of $\eta(1)$ when $\beta = 0.142$, $\gamma = 18.9$, $\alpha = 0.0018$, and a) $U = 6.6$, b) $U = 7.65$, c) $U = 8.94$, d) $U = 11.1$.

From the comparative studies conducted in this section, it can be concluded that the present quaternion-based model is capable of capturing the linear and nonlinear geometrically exact dynamics of hanging cantilevered pipe conveying fluid. As reported by different researchers, a critical issue with the exact geometrically rotation-based model is its high computational cost [27, 30, 32]. Although the quaternion-based model contains more equations with stronger nonlinearity, its computational cost is remarkably reduced compared to that of the rotation-based model. This is due to the fact that the Galerkin discretization method-based coefficients of the quaternion-based model are time-independent, unlike those of the rotation-based model. However, the critical issues with the solution strategy for the ordinary differential algebraic equations that arise from the quaternion-based model in this study are its accuracy for high values of dimensionless flow velocity and its solvability when it is needed to involve more approximation functions to obtain the accurate dynamic behavior of system. These challenging issues may be solved by using a numerical strategy that is capable of solving the differential-algebraic equations with the original constraint equation with index three rather than the reduced order one with index one or using the constraint stabilization techniques [43] for the differential-algebraic equations with reduced order constraint equation.

## 5. Conclusion

In the framework of this work, a new mathematical formulation for the geometrically exact analysis of cantilevered pipes conveying fluid is proposed. The linear and nonlinear quaternion-based formulations are derived by the implementation of the extended version of Hamilton's principle. The linear and nonlinear geometrically exact rotation-based formulations are recovered from the corresponding quaternion-based models. The new mathematical formulation contains two integro-differential equations and a constraint equation with index three. To be able to solve the problem at hand by using ode15s in MATLAB software, the index of the constraint equation is reduced to 1; however, the original constraint equation is studied after the solution. The ordinary differential algebraic equations that arise from discretizing the integro-partial differential-algebraic equations using the Galerkin technique are numerically solved. Different examples are then solved and compared with those reported in the existing literature to show the newly developed model is capable of successfully predicting the critical flow velocity at which the system undergoes flutter instability and simulating the nonlinear geometrically exact self-excited oscillations of the system in the post-flutter region. The comparative studies for the nonlinear dynamics include the time history, bifurcation diagrams, phase planes, and undeformed configurations over a steady-state oscillation of the system. The corresponding responses for the quaternion elements are also presented, and the original constraint equation is studied. The results show that the current model finds the solution much faster than the rotation-based model. However, in the numerical solution used in this study, there exist some difficulties. The first one is that the original constraint equation does not satisfy well when the dimensionless flow velocities are high, and the second one is the solvability problem for the cases in which more approximation functions are needed. One may solve these difficulties using the constraint stabilization techniques or applying the original constraint equation instead of the reduced-order one. These challenging issues can be considered for future study. Additionally, due to the gimbal lock of the three-dimensional rotation-based model, developing a three-dimensional quaternion-based formulation can be taken as another topic for future study.

## Appendix A

The coefficients related to the stiffness matrices due to the potential energy of pipe, the stiffness matrices due to the gravity, the mass matrices, the gyroscopic damping matrices, the structural damping matrices, the stiffness matrices to the fluid flow, the stiffness matrices due to the Lagrangian multiplier function, and the matrices due to the constraint equation, respectively, are as follows

$$K_{ijkl}^{11} = 4\int_0^1 \left( \phi_i \phi_j'' \psi_k \psi_l - \phi_i \phi_j \psi_k \psi_l'' - 2\phi_i \phi_j \psi_k' \psi_l' + 2\phi_i \phi_j' \psi_k \psi_l' \right) ds, \quad {}^1K_{ijk}^{11} = -4\left( \phi_i \psi_j \psi_k'' + 2\phi_i \psi_j' \psi_k' \right)$$
$$K_{ijkl}^{21} = 4\int_0^1 \left( \psi_i \phi_j \phi_k \psi_l'' - \psi_i \phi_j \phi_k'' \psi_l - 2\psi_i \phi_j' \phi_k' \psi_l + 2\psi_i \phi_j \phi_k' \psi_l' \right) ds, \quad (A1)$$
$${}^1K_{ijk}^{21} = 4\int_0^1 \left( 2\psi_i \phi_j \psi_k'' - \psi_i \phi_k'' \psi_l + 2\psi_i \phi_k' \psi' \right) ds, \quad {}^2K_{ij}^{21} = 4\int_0^1 \left( \psi_i \psi_j'' \right) ds,$$

$$K_{ij}^{12} = \gamma \int_0^1 2\phi_i \phi_j (1-s) ds, \quad {}^1K_i^{12} = \gamma \int_0^1 2\phi_i (1-s) ds,$$
$$K_{ij}^{22} = -\gamma \int_0^1 2\psi_i \psi_j (1-s) ds \quad (A2)$$

$$M_{ijkl}^{11} = M_{ijkl}^{12} = \int_0^1 2\phi_i\phi_j \int_1^s \int_0^{\hat{s}} 2(\phi_k\phi_l)\,d\bar{s}d\hat{s}ds, \quad M_{ijkl}^{13} = M_{ijkl}^{14} = -\int_0^1 2\phi_i\phi_j \int_1^s \int_0^{\hat{s}} 2\psi_k\psi_l\,d\bar{s}d\hat{s}ds,$$

$$M_{ijkl}^{15} = M_{ijkl}^{16} = \int_0^1 2\phi_i\psi_j \int_1^s \int_0^{\hat{s}} 2(\phi_k\psi_l)\,d\bar{s}d\hat{s}ds, \quad M_{ijkl}^{17} = \int_0^1 2\phi_i\psi_j \int_1^s \int_0^{\hat{s}} 2(\psi_k\phi_l)\,d\bar{s}d\hat{s}ds,$$

$$^1M_{ijk}^{11} = \int_0^1 2\phi_i \int_1^s \int_0^{\hat{s}} 2(\phi_j\phi_k)\,d\bar{s}d\hat{s}ds + \int_0^1 2\phi_i\phi_j \int_1^s \int_0^{\hat{s}} 2(\phi_k)\,d\bar{s}d\hat{s}ds, \quad ^2M_{ij}^{11} = \int_0^1 2\phi_i \int_1^s \int_0^{\hat{s}} 2(\phi_j)\,d\bar{s}d\hat{s}ds,$$

$$^1M_{ijk}^{12} = \int_0^1 2\phi_i \int_1^s \int_0^{\hat{s}} 2(\phi_j\phi_k)\,d\bar{s}d\hat{s}ds, \quad ^1M_{ijk}^{13} = ^1M_{ijk}^{14} = -\int_0^1 2\phi_i \int_1^s \int_0^{\hat{s}} 2\psi_j\psi_k\,d\bar{s}d\hat{s}ds,$$

$$^1M_{ijk}^{15} = \int_0^1 2\phi_i\psi_j \int_1^s \int_0^{\hat{s}} 2(\psi_k)\,d\bar{s}d\hat{s}ds,$$

$$M_{ijkl}^{21} = M_{ijkl}^{22} = -\int_0^1 2\psi_i\psi_j \int_1^s \int_0^{\hat{s}} 2\phi_k\phi_l\,d\bar{s}d\hat{s}ds, \quad M_{ijkl}^{23} = M_{ijkl}^{24} = \int_0^1 2\psi_i\psi_j \int_1^s \int_0^{\hat{s}} 2\psi_k\psi_l\,d\bar{s}d\hat{s}ds,$$

$$M_{ijkl}^{25} = M_{ijkl}^{26} = \int_0^1 2\psi_i\phi_j \int_1^s \int_0^{\hat{s}} 2(\phi_k\psi_l)\,d\bar{s}d\hat{s}ds, \quad M_{ijkl}^{27} = \int_0^1 2\psi_i\phi_j \int_1^s \int_0^{\hat{s}} 2(\psi_k\phi_l)\,d\bar{s}d\hat{s}ds,$$

$$^1M_{ijk}^{21} = -\int_0^1 2\psi_i\psi_j \int_1^s \int_0^{\hat{s}} 2\phi_k\,d\bar{s}d\hat{s}ds, \quad ^2M_{ij}^{25} = \int_0^1 2\psi_i \int_1^s \int_0^{\hat{s}} 2(\psi_l)\,d\bar{s}d\hat{s}ds,$$

$$^1M_{ijk}^{25} = \int_0^1 2\psi_i \int_1^s \int_0^{\hat{s}} 2(\phi_j\psi_l)\,d\bar{s}d\hat{s}ds + \int_0^1 2\psi_i\phi_j \int_1^s \int_0^{\hat{s}} 2(\psi_k)\,d\bar{s}d\hat{s}ds,$$

$$^1M_{ijk}^{26} = \int_0^1 2\psi_i \int_1^s \int_0^{\hat{s}} 2(\phi_j\psi_k)\,d\bar{s}d\hat{s}ds, \quad ^1M_{ijk}^{27} = \int_0^1 2\psi_i \int_1^s \int_0^{\hat{s}} 2(\psi_j\phi_k)\,d\bar{s}d\hat{s}ds,$$

(A3)

$$G_{ijkl}^{11} = 2U\sqrt{\beta}\int_0^1 2\phi_i\phi_j \int_1^s 2(\phi_k\phi_l)\,d\hat{s}ds, \quad G_{ijkl}^{12} = -2U\sqrt{\beta}\int_0^1 2\phi_i\phi_j \int_1^s 2\psi_k\psi_l\,d\hat{s}ds,$$

$$^2G_{ij}^{11} = 2U\sqrt{\beta}\int_0^1 2\phi_i \int_1^s 2(\phi_j)\,d\hat{s}ds, \quad ^1G_{ijk}^{11} = 2U\sqrt{\beta}\left(\int_0^1 2\phi_i \int_1^s 2(\phi_j\phi_k)\,d\hat{s}ds + \int_0^1 2\phi_i\phi_j \int_1^s 2(\phi_k)\,d\hat{s}ds\right),$$

$$^1G_{ijk}^{12} = -2U\sqrt{\beta}\int_0^1 2\phi_i \int_1^s 2\psi_j\psi_k\,d\hat{s}ds, \quad G_{ijkl}^{13} = G_{ijkl}^{14} = 2U\sqrt{\beta}\int_0^1 2\phi_i\psi_j \int_1^s 2\psi_k\phi_l\,d\hat{s}ds,$$

$$^1G_{ijk}^{14} = 2U\sqrt{\beta}\int_0^1 2\phi_i\psi_j \int_1^s 2\psi_k\,d\hat{s}ds,$$

$$G_{ijkl}^{21} = -2U\sqrt{\beta}\int_0^1 2\psi_i\psi_j \int_1^s 2\phi_k\phi_l\,d\hat{s}ds, \quad G_{ijkl}^{22} = 2U\sqrt{\beta}\int_0^1 2\psi_i\psi_j \int_1^s 2\psi_k\psi_l\,d\hat{s}ds,$$

$$^1G_{ijk}^{22} = -2U\sqrt{\beta}\int_0^1 2\psi_i\psi_j \int_1^s 2\phi_k\,d\hat{s}ds, \quad G_{ijkl}^{23} = G_{ijkl}^{24} = 2U\sqrt{\beta}\int_0^1 2\psi_i\phi_j \int_1^s 2\psi_k\phi_l\,d\hat{s}ds,$$

$$^1G_{ijk}^{23} = 2U\sqrt{\beta}\int_0^1 2\psi_i \int_1^s 2\psi_j\phi_k\,d\hat{s}ds, \quad ^2G_{ij}^{24} = 2U\sqrt{\beta}\int_0^1 2\psi_i \int_1^s 2\psi_j\,d\hat{s}ds,$$

$$^1G_{ijk}^{24} = 2U\sqrt{\beta}\left(\int_0^1 2\psi_i\phi_j \int_1^s 2\psi_k\,d\hat{s}ds + \int_0^1 2\psi_i \int_1^s 2\phi_j\psi_k\,d\hat{s}ds\right),$$

(A4)

$$C^{11}_{ijkl} = 4\alpha \int_0^1 \left(2\phi_i\psi'_j\psi_k\phi'_l - 2\phi_i\psi'_j\psi'_k\phi_l + \phi_i\psi_j\psi'_k\phi' + \phi_i\psi_j\psi_k\phi'' - \phi_i\psi_j\psi''_k\phi_l - \phi_i\psi_j\psi'_k\phi'_l\right)ds,$$

$$C^{12}_{ijkl} = 4\alpha \int_0^1 \left(2\phi_i\psi'_j\phi'_k\psi_l - 2\phi_i\psi'_j\phi_k\psi' + \phi_i\psi_j\phi'_l\psi' + \phi_i\psi_j\phi''_k\psi_l - \phi_i\psi_j\phi_k\psi''_l - \phi_i\psi_j\phi'\psi'_l\right)ds,$$

$$C^1_{12} = -8\alpha \int_0^1 \left(\phi_i\psi'_j\psi'_k - \phi_i\psi'_j\psi'_k\right)ds,$$

$$C^{21}_{ijkl} = 4\alpha \int_0^1 \left(-2\psi_i\phi'_j\psi_k\phi'_l + 2\psi_i\phi'_j\psi'_k\phi_l - \psi_i\phi_j\psi'_k\phi'_l - \psi_i\phi_j\psi_k\phi'' + \psi_i\phi_j\psi''_k\psi_l + \psi_i\phi_j\psi'_k\phi'\right)ds,$$

(A5)

$$C^{22}_{ijkl} = 4\alpha \int_0^1 \left(-2\psi_i\phi'_j\phi'_k\psi_l + 2\psi_i\phi'_j\phi_k\psi'_l - \psi_i\phi_j\phi'\psi'_l - \psi_i\phi_j\phi''_k\psi_l + \psi_i\phi_j\phi_k\psi'' + \psi_i\phi_j\phi'_k\psi'\right)ds,$$

$$^1C^{21}_{ijk} = 4\alpha \int_0^1 \left(-\psi_i\psi'_j\phi'_k - \psi_i\psi_j\phi''_k + \psi_i\psi''_j\psi_k + \psi_i\psi'_j\phi'\right)ds,$$

$$^1C^{22}_{ijk} = 4\alpha \int_0^1 \left(2\psi_i\phi'_j\psi' - \psi_i\phi'_j\psi' - \psi_i\phi''_j\psi_k + 2\psi_i\phi_j\psi''_k + \psi_i\phi'_j\psi'_k\right)ds,$$

$$^2C^{22}_{ij} = 4\alpha \int_0^1 \left(\psi_i\psi''_k\right)ds,$$

$$K^{13}_{ijkl} = -U^2 \int_0^1 2\phi_i\phi_j\left(\phi_k(1)\phi_l(1)\right)ds, \quad ^2K^{13}_{ij} = -U^2 \int_0^1 2\left(\phi_i\phi_j + 2\phi_i\phi_j(1)\right)ds,$$

$$^1K^{13}_{ijk} = -U^2 \int_0^1 2\left(2\phi_i\phi_j\phi_k(1) + \phi_i\phi_j(1)\phi_k(1)\right)ds, \quad ^3K^{13}_i = -U^2 \int_0^1 2\phi_i ds,$$

$$K^{14}_{ijkl} = U^2 \int_0^1 2\phi_i\phi_j\psi_k(1)\psi_l(1)ds, \quad K^{15}_{ijkl} = -U^2 \int_0^1 2\phi_i\psi_j\left(2\phi_k(1)\psi_l(1)\right)ds,$$

$$^1K^{14}_{ijk} = U^2 \int_0^1 2\phi_i\psi_j(1)\psi_k(1)ds, \quad ^1K^{15}_{ijk} = -U^2 \int_0^1 2\phi_i\psi_j\left(2\psi_k(1)\right)ds$$

(A6)

$$K^{23}_{ijkl} = U^2 \int_0^1 2\psi_i\psi_j\left(\phi_k(1)\phi_l(1)\right)ds,$$

$$^2K^{23}_{ij} = U^2 \int_0^1 2\psi_i\psi_j ds, \quad ^1K^{23}_{ijk} = U^2 \int_0^1 2\psi_i\psi_j\left(2\phi_k(1)\right)ds,$$

$$K^{24}_{ijkl} = -U^2 \int_0^1 2\psi_i\psi_j\psi_k(1)\psi_l(1)ds, \quad K^{25}_{ijkl} = -U^2 \int_0^1 2\psi_i\phi_j\left(2\phi_k(1)\psi_l(1)\right)ds$$

$$^2K^{25}_{ij} = -U^2 \int_0^1 2\psi_i\left(2\psi_j(1)\right)ds, \quad ^1K^{25}_{ijk} = -U^2 \left(\int_0^1 2\psi_i\phi_j\left(2\psi_k(1)\right)ds + \int_0^1 2\psi_i\left(2\phi_j(1)\psi_k(1)\right)ds\right),$$

$$K^{16}_{ijk} = -\int_0^1 2\phi_i\phi_j\chi_k ds, \quad ^1K^{16}_{ij} = -\int_0^1 2\phi_i\chi_j ds, \quad K^{26}_{ijk} = -\int_0^1 2\psi_i\psi_j\chi_k ds,$$

(A7)

$$D^{31}_{ijk} = D^{33}_{ijk} = \int_0^1 \chi_i\phi_j\phi_k ds, \quad ^1D^{33}_{ij} = \int_0^1 \chi_i\phi_j ds, \quad D^{32}_{ijk} = D^{34}_{ijk} = \int_0^1 \chi_i\psi_j\psi_k ds.$$

(A8)